\documentclass[twocolumn,floatfix,superscriptaddress,aps,prl,noeprint]{revtex4-2}
 
\usepackage{graphicx} 
\usepackage{mathtools,color,float,amsmath}\usepackage[a4paper]{geometry}
\usepackage{xcolor}

\usepackage{braket}
\usepackage{amsfonts,amssymb,amsmath,mathrsfs}
\usepackage{hyperref}
\hypersetup{colorlinks=true,linkcolor=blue,citecolor=blue}
\usepackage{natbib}
\usepackage[utf8]{inputenc}
\usepackage[section]{placeins}

\begin{document}
\title{Deterministic single-photon source over the terahertz regime}

\date{\today}
\author{Caspar Groiseau}
\email{groiseau@chalmers.se}
\affiliation{Department of Microtechnology and Nanoscience, Chalmers University of Technology, 412 96 Gothenburg, Sweden}	
\affiliation{Departamento de F\'isica Teórica de la Materia Condensada and Condensed 
Matter Physics Center (IFIMAC), Universidad Autónoma de Madrid, 28049 Madrid, Spain}

\author{Miguel Á. Martínez-García}
\affiliation{Departamento de F\'isica Teórica de la Materia Condensada and Condensed Matter Physics Center (IFIMAC), Universidad Autónoma de Madrid, 28049 Madrid, Spain}

\author{Diego Martín-Cano}
\email{diego.martin.cano@uam.es}
\affiliation{Departamento de F\'isica Teórica de la Materia Condensada and Condensed Matter Physics Center (IFIMAC), Universidad Autónoma de Madrid, 28049 Madrid, Spain}

\author{Carlos Sánchez Muñoz}
\email{carlos.sanchez@iff.csic.es}
\affiliation{Institute of Fundamental Physics IFF-CSIC, Calle Serrano 113b, 28006, Madrid, Spain}

\begin{abstract}
The terahertz (THz) regime still lacks a source capable of emitting single photons deterministically. 
Here, we propose an on-demand THz single-photon source, triggered by a sequence of coherent optical pulses used to dress a polar quantum emitter. 
The permanent dipole moment of the emitter activates THz transitions within its Rabi-split doublets, which are enhanced through resonant coupling to a  THz cavity. 
 We identify a challenge unique to this platform: spontaneous emission of visible photons competes with the THz process, degrading performance in two distinct ways depending on the sign of the drive detuning---either increasing the vacuum component, reducing brightness, or increasing multiphoton contributions, reducing purity.
We demonstrate how to overcome those challenges through optimized pulse areas and a sufficiently high Purcell rate. 
A hybrid resonator-nanoparticle design for a cavity satisfying the latter requirement is presented, yielding efficiencies of 65-92\% and purities of 88-100\%, comparable to the best visible single-photon sources, while providing broad tunability across the THz range. Finally, we exploit the visible emission itself, using its detection to fully reconstruct the THz photon-number distribution and herald a single THz photon, boosting efficiency to 90-95\%. All these capabilities illustrate the new technological opportunities unlocked by the unique integration of terahertz and visible frequencies.
\end{abstract}
\maketitle

\emph{Introduction---}Single photon emitters (SPE) are essential building blocks for the generation of non-classical correlations involved in light-based quantum technologies~\cite{walmsley2015,aharonovich2016,gonzalez-tudela2024}. To perform useful 
quantum protocols, SPEs are required to fulfill 
bright and efficient emission properties, and
the ability to trigger on demand indistinguishable single photons with a well-defined spatial-temporal mode. These demands have led to a contest to integrate a plethora of different solid-state emitters in various nanophotonic platforms, including quantum dots~\cite{lodahl2015}, single defects~\cite{aharonovich2016}, superconducting qubits~\cite{blais2021} and molecules~\cite{toninelli2021}, with the prospects of becoming standard sources for quantum optical technologies.

While the race for SPEs began long ago for visible and microwave wavelengths, it has just begun for Terahertz (THz) radiation (0.1-70 THz) \cite{tonouchi2007}. The THz is among the latest exploited gaps in the electromagnetic spectrum, with broad applications in communications, biology,  medicine, spectroscopy and materials characterization~\cite{tonouchi2007,zhang2017}. 
The historical delay in the development of classical THz technology, together with the absence of bright and efficient photonic transitions at THz frequencies~\cite{todorov2024}---naturally occurring in molecules, superconducting junctions, or semiconductors---have prevented any experimental findings of THz SPEs. Only few proposals of heralded THz photon sources exist, such as spontaneous parametric down-conversion experiments~\cite{Leontyev2021} and polariton optomechanical mechanisms in the mid-infrared~\cite{delPino2016, Iles-Smith2025, Shishkov2025}. Developing such THz SPEs would set an alternative quantum optical platform offering a compromise between the challenges of the visible (nanometric tolerance requirements) and microwave spectrum (scalability, mK cooling).

A promising approach for generating THz radiation involves leveraging the interaction between cavities and laser-dressed electronic transitions in polar systems~\cite{kibis2009,savenko2012,shammah2014,chestnov2017a,deliberato2018,pompe2023}. Permanent dipole moments in asymmetric emitters enable nonlinear radiative transitions between laser-dressed states of the same Rabi doublet, which can achieve THz frequencies. In a recent work~\cite{groiseau2024}, we have shown that Purcell enhancement of this mechanism by THz cavities allows one to generate a continuous THz SPE, which bypasses the large ratio between dephasing and transition rates that hinder conventional optical dipole transitions at THz frequencies~\cite{cole2001}. 
However, continuous SPEs do not allow deterministic timing of photon emission events and results in emission over steady-state mixtures within finite detection windows, limiting indistinguishability
and compatibility with synchronized quantum protocols.

\begin{figure*}
 \includegraphics[width=0.9\textwidth]{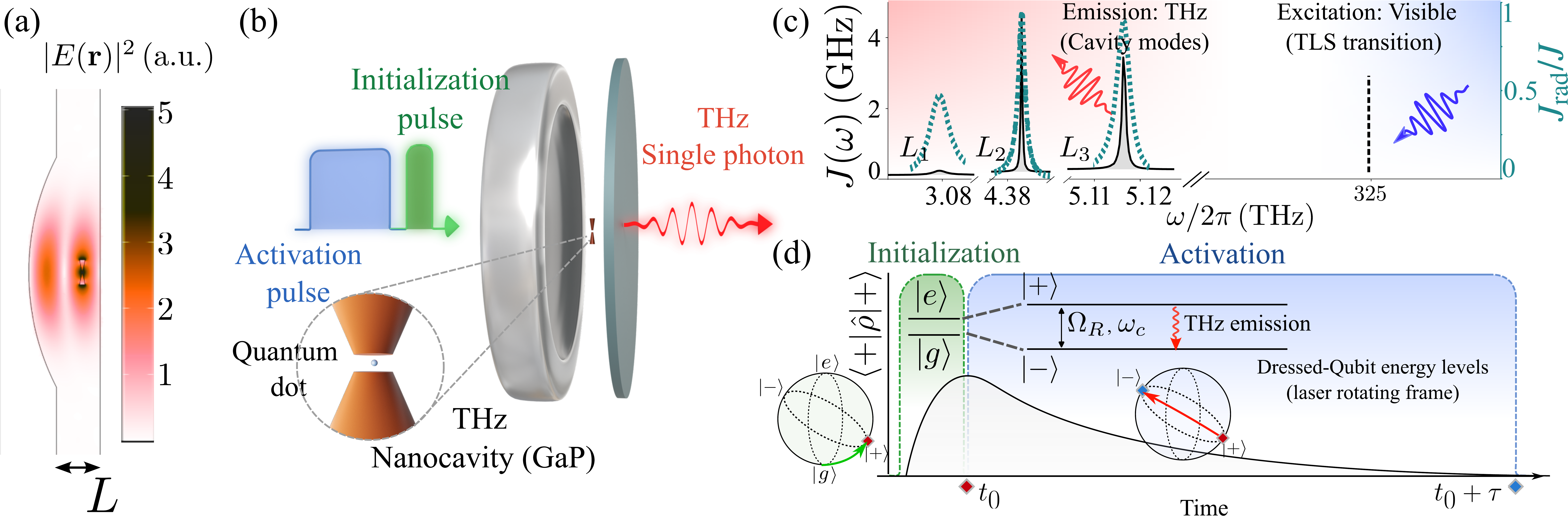}
    \caption{(a) Electric field cross-section simulation and sketch (b) of a potential experimental implementation of a deterministic single photon THz source consisting of a quantum emitter held in a hybrid nanocavity made up by two GaP nanocones (length $11.07~\mu\text{m}$, tip radius $0.39~\mu\text{m}$, bottom radius $2.37~\mu\text{m}$) separated by 50nm, both embedded within a Fabry-Pérot resonator which is triggered by the initialization and activation pulse protocols in the visible (green and blue areas). (c) Simulations of the spectral density $J$  and its radiative part $J_\text{rad}$ of three different frequency-detuned hybrid cavities with lengths $\{L_1,L_2,L_3\} =    \{108.9~\mu\text{m}, 74.7~\mu\text{m}, 63.4~\mu\text{m}\}$.  (d): Temporal evolution of the population of $|+\rangle$ during the first two pulses, including trajectory on the Bloch sphere. 
    }
    \label{fig1}  
\end{figure*}

Here, we develop a full scheme for on-demand, optically-heralded generation of single photons in the THz regime with high purity, efficiency, and indistinguishability. 
We realize this by combining a tailored pulse protocol with a feasible hybrid-cavity design [see Fig.~\ref{fig1}(a,b)]. 
While competing visible-photon emission can degrade the output purity, its impact can be minimized through optimized cavity and pulse design. At the same time, these visible photons enable heralding and complete tomographic reconstruction of the photon-number distribution of the emitted THz state using fully optical detection.

\emph{System and model---} We use a single optical polar quantum emitter described as a two-level system (TLS) with ground state $|g\rangle$ and  excited state $|e\rangle$, separated by the optical transition frequency $\omega_0$. The TLS interacts with a single THz cavity mode represented by the annihilation operator $\hat a$, characterized by the THz frequency $\omega_c$ and the associated field $\textbf{E}_c$. 
In our proposal, this mode is realized through a hybrid cavity structure consisting of an Fabry-Pérot resonator coupled to a THz nanocavity formed by two separated GaP nanocones, as illustrated in Fig.~\ref{fig1}(a,b). These structures, as well as the required emitter positioning, are within the capabilities of current fabrication technologies \cite{sapienza2015,liu2021}, the cones being  supported either by tips or by ultra-low-refractive-index substrates, which have already been successfully employed in THz microcavity implementations \cite{kim2023,zhang2004}. 
We perform full electromagnetic dipole simulations~\cite{zotero-492} to estimate the photonic spectral local density at the center of the nanocavity at THz frequencies, see Fig.~\ref{fig1}(c).  
 This hybrid architecture enables broadband tunability of the THz emission by leveraging the external control of the cavity length $L$---see different mode spectral densities in Fig.~\ref{fig1}(c) for varying $L$---, together with the flat response of GaP at frequencies well below phonon polariton resonances~\cite{foteinopoulou2019}. These cavity modes enable a $10^6$-fold emission enhancement with respect to free space and high radiative efficiency, due to the high refractive index of the semiconductor material \cite{foteinopoulou2019} in combination with the finesse of the Fabry-Pérot resonator~\cite{gurlek2018}. Finally, the TLS is driven by an optical laser field $\textbf{E}_L$ with frequency $\omega_L$. 

This system is described through the Hamiltonian ($\hbar=1$ henceforth):
$    \hat H={\omega_0}\hat\sigma_z/2+\omega_c\hat a^\dagger \hat a+\hat{ \textbf{d}}\cdot\textbf{E}_c(\hat a+\hat a^\dagger)+\hat{ \textbf{d}}\cdot\textbf{E}_L\cos(\omega_L t),$
where the dipole operator $\hat{ \textbf{d}}$ is defined as $\hat{ \textbf{d}} = \textbf{d}_{ee}(1+\hat\sigma_z)/2 + \textbf{d}_{ge}(\hat\sigma_+ + \hat\sigma_-)$, $\hat\sigma_{\pm}$ being the raising and lowering operators of the TLS. The term   $\propto \textbf{d}_{ee}$ accounts for the permanent dipole component due to the asymmetric charge distribution of the polar emitter. After moving to the laser rotating frame and eliminating all terms oscillating with visible frequencies~\cite{groiseau2024}, the optical coherent drive yields two dressed eigenstates $|\pm\rangle$ within the TLS-laser system [see Fig.~\ref{fig1}(d)], 
split by the generalized Rabi frequency $\Omega_R = \sqrt{\Delta^2 + \Omega^2}$, where $\Delta = \omega_L - \omega_0$ represents the laser detuning, and $\Omega = \textbf{d}_{ge}\cdot\textbf{E}_L$ denotes the driving amplitude~\cite{chestnov2017a,sanchezmunoz2018}. These dressed states, $|+\rangle = s|e\rangle + c|g\rangle$ and $|-\rangle = s|g\rangle - c|e\rangle$, are defined by $s \equiv \sin\theta$, $c \equiv \cos\theta$, where $\theta\equiv\arctan\left[h^{\operatorname{sgn}(\Delta)}\right]\in[0,\frac{\pi}{4}]$, and where $h\equiv\frac{\Omega_R-|\Delta|}{\Omega} \in [0,1]$, which we term dressing ratio, characterizes the crossover from the undressed regime ($h=0$) to the fully dressed regime of resonant driving ($h=1$).
The operators $\hat\sigma_{\pm,z}$ are straightforwardly expressed using the Pauli matrices of the dressed-state basis, denoted as $\hat\zeta_{\pm}\equiv|\pm\rangle\langle\mp|$, $\hat\zeta_{z}\equiv|+\rangle\langle+|-|-\rangle\langle-|$
\footnote{$\hat\sigma_{\pm} = cs\hat\zeta_z + s^2\hat\zeta_\pm - c^2\hat\zeta_\mp$ and $\hat\sigma_{z} = (s^2-c^2)\hat\zeta_z-2cs(\hat\zeta_++\hat\zeta_-)$}. In the dressed basis, the resulting Hamiltonian reads $\hat H=\frac{\Omega_R}{2}\hat \zeta_z+\omega_c\hat a^\dagger\hat a-2cs\chi(\hat a\hat\zeta_++\hat a^\dagger\hat\zeta_-)
    -2cs\chi(\hat a\hat\zeta_-+\hat a^\dagger\hat\zeta_+)+\chi(\hat a+\hat a^\dagger)[1+(s^2-c^2)\hat\zeta_z]$, where $\chi=\textbf{d}_{ee}\cdot\textbf{E}_c/2$ is the coupling rate between the TLS and the THz cavity.

Additionally, we describe cavity decay and TLS spontaneous emission at optical frequencies with rates $\kappa$ and $\gamma$, respectively. The full open quantum system used in our simulations is thus given by the Master equation $ \mathcal{L}{\hat\rho}\equiv \partial_t \hat\rho =-i[\hat H,\hat \rho]+\frac{\kappa}{2}\mathcal{D}( \hat X)\hat\rho
+\frac{1}{2}\mathcal{D}(-\sqrt{\gamma_+}\hat\zeta_++\sqrt{\gamma_-}\hat\zeta_-+\sqrt{\gamma_z}\hat\zeta_z)\hat\rho$, where $\gamma_-=\gamma s^4$, $\gamma_+=\gamma c^4$,  $\gamma_z=\gamma c^2s^2$, $\mathcal{L}$ is the Liouvillian superoperator and $\mathcal{D}(\hat O)\hat\rho=2\hat O\hat \rho \hat O^\dagger-\hat O^\dagger \hat O\hat \rho-\hat \rho \hat O^\dagger \hat O$ is the Lindblad superoperator. 
The cavity decay is described by the operator $\hat X=\sum_{j,k>j}\sqrt{\frac{\omega_k-\omega_j}{\omega_c}}\langle j|(\hat a+\hat  a^\dagger)|k\rangle|j\rangle\langle k|$, which accounts for the counter-rotating terms required beyond the secular approximation~\cite{ridolfo2012,groiseau2024}. Here, $|k\rangle$ is the $k$-the eigenstate with energy $\omega_k$.

Under the resonant condition $\omega_c = \Omega_R$, and provided $\chi\ll\omega_c$, we can eliminate terms oscillating at THz frequencies and get a Jaynes-Cummings Hamiltonian
$\hat H_\text{JC}=g(\hat a\hat\zeta_++\hat a^\dagger\hat\zeta_-)$~\cite{chestnov2017a,groiseau2024}, where we have defined the effective coupling strength $g=-2cs\chi$.
Two further approximations will allow us to obtain analytical results:
(i) First, we assume that we are in the far-off-resonant regime ($|\Delta|\gg\Omega$). This implies that $h\ll1$, allowing us to set $\gamma_{\mp}=\gamma_z\approx 0$ and $\gamma_\pm=\gamma$, where from now on we refer the upper and lower sub-indexes to $\Delta>0$  and $\Delta<0$, respectively.
(ii) Second, we assume and irreversible, dissipative coupling to the cavity, which can be eliminated under the condition $\kappa\gg\gamma,g$. Using standard adiabatic elimination techniques \cite{gonzalez-ballestero2024}, we obtain 
$\mathcal{L}_\text{adb}\,{\hat\rho}\equiv \frac{\gamma_\pm}{2}\mathcal{D}(\hat\zeta_\pm)\hat\rho+\frac{\Gamma}{2}\mathcal{D}(\hat\zeta_-)\hat\rho
                $,  
where $\Gamma\equiv4g^2/\kappa$ is the effective Purcell rate associated with the characteristic time scale for the emission of a THz photon through the cavity~\footnote{In this derivation we discarded the coherent part which only consists in an energy shift inconsequential for $P$ since $[\hat \zeta_+\hat\zeta_-,|\pm\rangle\langle\pm|]=0$}. We are left with two competing processes: the desired decay through the cavity with rate $\Gamma$ and a detrimental incoherent pump/decay with rate $\gamma_\pm$. 

\begin{table}[t!]
\centering
\begin{tabular}{|l|c|c|c|c|c|c|}
\hline
 & $\omega_c/2\pi$ & $\chi/2\pi$ & $\kappa/2\pi$  & $C$ & $\tilde C$ \\
\hline
Cavity $L_1$ &  $3.08\ \text{THz}$ & $4\ \text{GHz}$ & $3.5\ \text{GHz}$  &  460   & 1.94  \\ 
\hline
Cavity $L_3$ &  $5.12\ \text{THz}$ & $12.5\ \text{GHz}$ & $0.97\ \text{GHz}$  &  16200  & 24.65  \\ 
\hline
\end{tabular}
\caption{Parameters for two cavity configurations obtained by full electromagnetic dipole simulations.}
\label{tab:parameters}
\end{table}

\emph{Protocol---}The deterministic protocol consists of two consecutive optical laser pulses [see Fig.~\ref{fig1}(b,d)] that are applied sequentially to the ground state of the TLS-cavity system, $\hat\rho (0)=|g\rangle\langle g|\otimes|0\rangle\langle0|_\text{THz}$ (both pulses have square shape). 
(i) First, a short initialization pulse prepares the system at time $t_0$ in a coherent superposition $\hat\rho(t_0) = |+\rangle\langle+| \otimes |0\rangle\langle0|_\text{THz}$, where $|+\rangle$ is the dressed excited state defined for a chosen angle $\theta$. This state is prepared by a rotation of angle $\theta$ around the $y$-axis, $\hat U_1 = \exp[i\hat\sigma_y \theta]$, which can be implemented using a drive with Hamiltonian $\hat H_1 = i\Omega_1(\hat\sigma - \hat\sigma^\dagger)$ applied for a duration $t_0 = \theta/\Omega_1$ (alternatively, any pulse shape with the same area and phase---e.g., a Gaussian---can be used).
(ii) Second, a long activation pulse $\hat H_d = \Omega(\hat\sigma + \hat\sigma^\dagger)/2$, of duration $\tau$, dresses the system, creating Rabi doublets with THz splittings which couple resonantly with the cavity, $\Omega_R=\omega_c$. This induces a Purcell-enhanced radiative decay from $|+\rangle$ to $|-\rangle$ with rate $\Gamma$ via the emission of a THz photon [see Fig.~\ref{fig1}(d)].

\emph{SPE figures of merit---} 
In an ideal deterministic SPE, the probability $P_1$ of emitting exactly one photon per excitation cycle---commonly referred to as the brightness or \emph{efficiency} of the source~\cite{senellart2017,esmann2024}---is equal to unity. We thus define $P_1(\tau)$ as the probability that the state $\hat\rho_0$ evolves over a time interval $\tau$ while emitting exactly one photon. This is given by $P_1(\tau)      =\int_0^\tau\text{Tr}(S_{\tau-t}\mathcal{K}S_t\hat\rho_0) dt$,
where $\mathcal{K}\hat\rho\equiv\kappa\hat X\hat\rho\hat X^\dagger$  is the superoperator describing the quantum collapse after photon detection, and $S_t=e^{(\mathcal{L}-\mathcal{K})t}$ is the conditional propagator for a time period $t$ conditioned to no photodetection.

A single-photon source is also characterized by the indistinguishability and the purity~\cite{senellart2017,esmann2024}. The indistinguishability $I$---formally defined in Sec.~I of the Supplemental Material (SM) \cite{supp}---quantifies to which degree different single-photon pulses share the same wave function, which is verified via Hong-Ou-Mandel interferometry. 
The purity $\Pi=1-g^{(2)}(0)$ informs about the likelihood of multiple photons being in the pulse, as established by the degree of quantum second-order coherence $g^{(2)}(0)$ measured in a Hanbury-Brown Twiss (HBT) setup. 
In this system, $I$ and $\Pi$ take essentially the same value.

\begin{figure}[t]
	\includegraphics[width=0.99\linewidth]{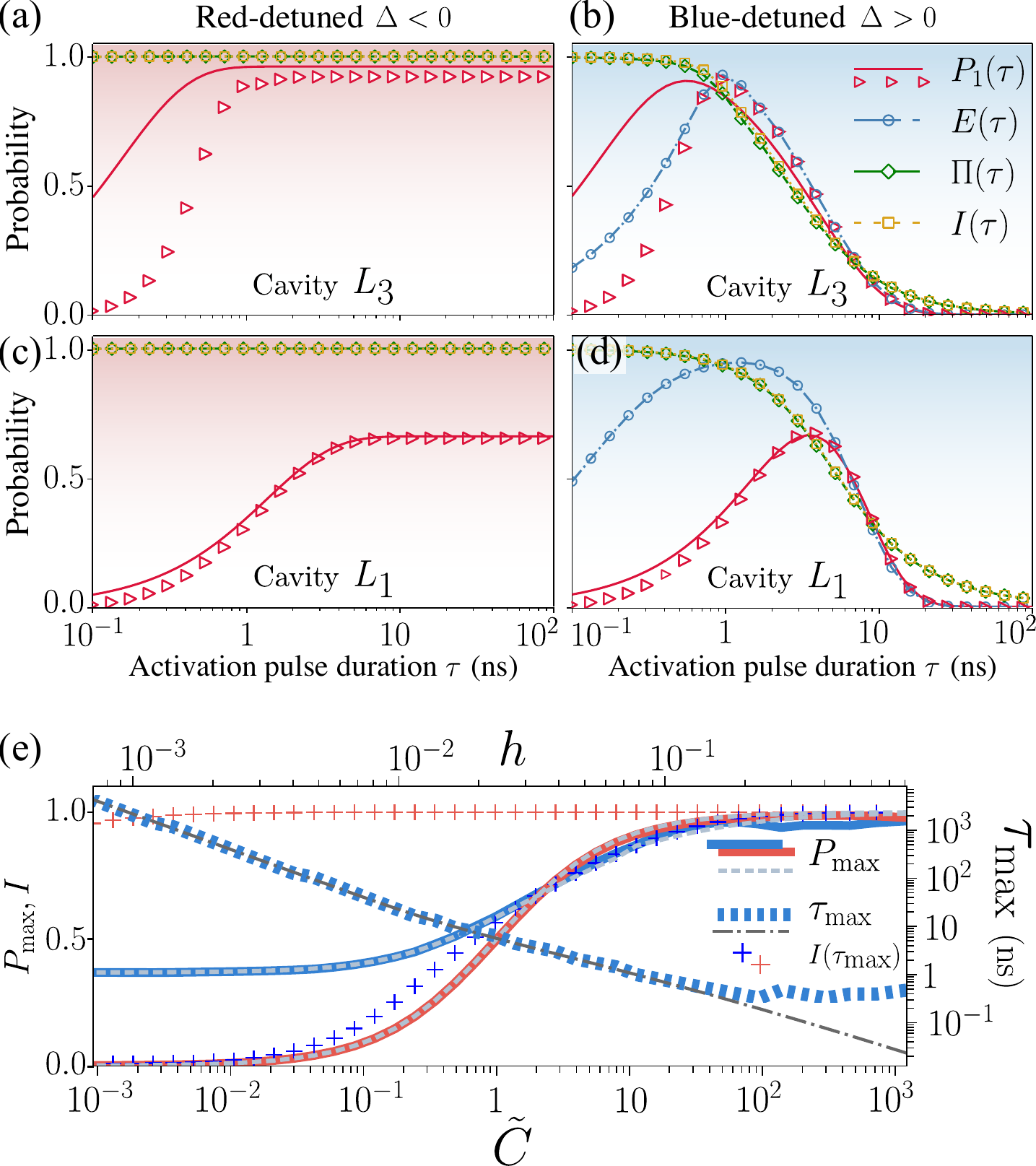}
     \caption{(a-d) $P_1(\tau)$ obtained numerically (solid red) and analytically via Eq.~\eqref{probability} (red triangles); heralding efficiency $E(\tau)$ (blue circles); purity $\Pi(\tau)$ (green diamonds) and indistinguishability $I(\tau)$ (yellow squares) versus activation pulse time $\tau$. 
     Results shown for cavity with length $L_3 =63.4\,\mu\text{m}$ (a-b) and 
    $L_1 =108.9\,\mu\text{m}$ (c-d).  
     Parameters: $\Omega/2\pi = 0.2\,\text{THz}$, corresponding to $h= 0.01955$ and $\tilde C = 24.65$ for cavity $L_3$, and $h= 0.0325$ and $\tilde C = 1.94$ for cavity $L_1$.
(e) Results for cavity length $L_1$ for the maximum single-photon detection probability $P_\text{max}$ (solid red and blue, for red and blue detuning, respectively); optimal activation pulse duration $\tau_\text{max}$ for the blue-detuned case (dashed blue), and their corresponding analytical estimates (dashed black for $P_\text{max}$ and dash-dotted black for $\tau_\text{max}$); and indistinguishability $I(\tau_\text{max})$ (blue and red crosses, for red and blue detuning, respectively), plotted as functions of the effective cooperativity  $\tilde C$, varied via the detuning $\Delta$, which effectively modifies the dressing ratio $h$ (upper axis). } 
    \label{fig:fig2}
\end{figure}

Fig.~\ref{fig:fig2}(a-d) presents exact results of efficiency $P_1$ versus activation pulse length $\tau$. The two columns correspond correspond to red- and blue-detuning respectively, with the same absolute value of detuning. The two rows correspond to
 two distinct cavity lengths ($L_1 \approx 108.9~\mu\text{m}$ and $L_3 \approx 63.4~\mu\text{m}$, see corresponding parameters in Table~\ref{tab:parameters}), highlighting the straightforward tunability of the source via simple modifications of the cavity geometry. 
These results  establish the crucial role of the drive detuning sign. A red-detuned drive stabilizes the maximum possible $P_1$ for any time $\tau \gg 1/\Gamma$ long enough to enable the cavity-mediated emission. In contrast, blue-detuning requires an optimal pulse duration: excessively long pulses degrade the purity by introducing higher-photon number components through effective repumping induced by spontaneous emission of visible photons at rate $\gamma$.

To obtain further insights, we now derive approximate analytical expressions from the effective TLS model given by $\mathcal{L}_\text{adb}$. 
 Starting with the initial state $\hat\rho_0=|+\rangle\langle+|$ set by the first pulse, we can straightforwardly compute $P_1(\tau)$ as
\begin{equation}  
\begin{split} 
         P_1(\tau) =\begin{cases}
			\frac{ e^{-\gamma \tau}-e^{-\Gamma \tau}[1+\gamma \tau(1-1/\tilde C)]}{(1-1/\tilde C)^2}, & \Delta>0\\
            \frac{\tilde C}{\tilde C+1}[1-e^{-\tau(\gamma+\Gamma )}], & \Delta<0.
		 \end{cases}
     \label{probability}
\end{split}
\end{equation}
Here, we introduced the effective cooperativity $\tilde C
\equiv 4g^2/[\kappa(\gamma_-+\gamma_+)]=4C/(h^2+h^{-2})\approx4h^2C$, a version of the standard cooperativity $C=4\chi^2/\kappa\gamma$, scaled by the factor $h$ to account for the dependence of the coupling rate on the dressing. 
In the case $\Delta>0$ and for $\tilde C \gg 1$, the  optimum pulse duration $\tau_\text{max}$ maximizing $P_1$ is given by $\tau_\text{max} \simeq \Gamma^{-1}\ln \tilde{C}$, yielding a maximum probability $P_\text{max}\simeq 1 - \ln \tilde{C}/\tilde{C}$. In this limit, the single-photon purity can be estimated to be $\Pi(\tau_\text{max})=1-2\ln\tilde C/\tilde C$. For $\Delta<0$, we can simply estimate $P_\text{max}\simeq \tilde C/(\tilde C+1)$, to which the system relaxes on a timescale $\tau_\text{rel}\sim(\Gamma+\gamma)^{-1}$, and $\Pi\simeq I\simeq 1$.

We see in Fig.~\ref{fig:fig2}(c-d) that Eq.~\eqref{probability} agrees well with the exact numerical results in a regime in which all of our assumptions are met. Interestingly, a configuration with cavity length $L_3$ [Fig.~\ref{fig:fig2}(a-b)] features a small decay rate $\kappa$ that does not meet the requirements of strong dissipation $\kappa \gg g$. Nevertheless, the source operates as an efficient SPE, with shorter $\tau_\text{max}$ and higher $P_\text{max}\approx 0.9$. This highlights that, although the conditions used to derive $\mathcal{L}_\text{adb}$ allow us to draw useful insights, they are not strictly necessary to achieve high-performance SPE operation.

Figure~\ref{fig:fig2}(e) shows the dependence of $P_\text{max}$ and $\tau_\text{max}$ as a function of $h$ for a cavity length $L_1$ (using the full model given by $\mathcal{L}$).
These results indicate that the source performance is governed by $\tilde C$, via the bare cooperativity $C$ and the dressing ratio $h$, suggesting that maximizing efficiency and minimizing $\tau_\text{max}$ benefits from working in the near-resonant regime $h\approx 1$ and large  $\tilde C$. However, near-resonant driving entails some complications that make this a challenging regime of operation, such as larger required values of $\Omega$ in the THz range and more important role of the decay terms $\gamma_\mp$ ignored in our analytics.  Furthermore, highly blue-detuned regimes provide the advantage of working with an initial state very close to the bare ground state, $\hat\rho_0 = |+\rangle\langle +|\approx |g\rangle\langle g|$, making the initialization pulse unnecessary (see SM Sec.~IV).
In conclusion, a compromise between large enough $P_\text{max}$ with small enough $h$ is desirable, such as the region $h\sim 0.1$ in the case shown in Fig.~\ref{fig:fig2}(e).

Regarding indistinguishability and purity, which are computed at $\tau_\text{max}$ ($\Delta >0$) and $ 3 \tau_\text{rel}$ ($\Delta <0$), our numerics show that, for $\Delta<0$, they always remains close to the maximum possible value, $I\approx \Pi \approx 1$. For $\Delta>0$, they  decreases with $\tilde C$ as the optimum pulse duration increases. In that case, a pulse duration shorter than the theoretical optimum is a more sensible choice, since it can maximize the heralding efficiency that we introduce below.

\emph{Optical heralding and tomography---} Up to this point, a blue-detuned configuration seemed to be the least favorable choice, as the repumping of the dressed TLS due to spontaneous emission in the visible leads to degradation of the quality of the SPE and requires pulse optimization. However, this configuration also allows one to use the detection of these visible photons to herald the THz SPE and even reconstruct its full photon-number distribution.

\begin{figure}[t]
	\centering \includegraphics[width=0.99\linewidth]{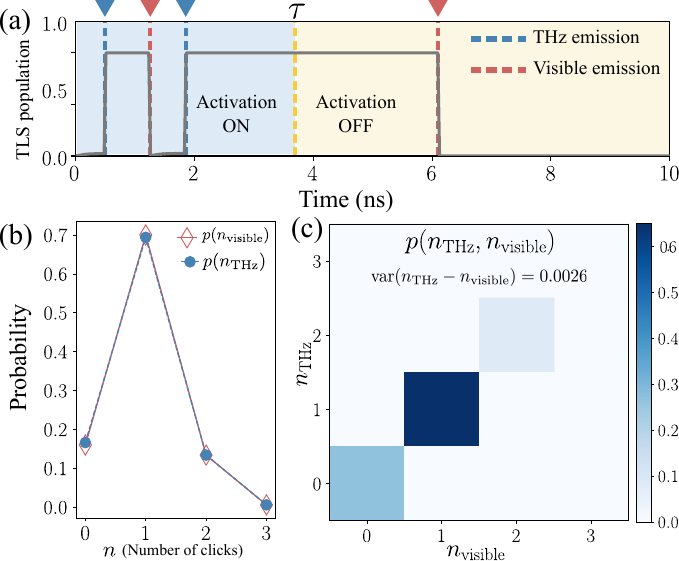}
    \caption{Optical heralding and tomography for cavity $L_1$, $h= 0.0325$, and activation pulse length $\tau = \tau_\text{max}$. (a) Single quantum trajectory in terms of the bare TLS population $\langle\hat\sigma_+\hat\sigma_-\rangle$ during the activation pulse (blue background), and the undriven evolution thereafter (yellow background). THz and visible photon emission events are marked, illustrating the one-to-one correspondence between the total numbers of clicks. (b) Probability distribution of the number of emitted photons in the THz ($n_\text{THz}$) and the visible ($n_\text{visible}$)---integrated over 30 ns after initialization pulse--- showing that both distributions match (due to the large detuning used, no heralding pulse is necessary). 
    (c) Joint probability distribution $p(n_\text{THz},n_\text{visible})$, showing close to perfect correlation between both quantities as summarized by the negligible  $\text{var}(n_\text{THz}-n_\text{visible})$.
   } 
    \label{fig:fig3-heralding}
\end{figure}

The key observation is that the state of the dressed emitter must be $|-\rangle$ following the emission of a THz photon, which for large blue detuning ($h\ll 1$ and $\Delta>0$),  approximates the bare excited state $|-\rangle \approx |e\rangle$.  This state can subsequently relax via spontaneous emission of a visible photon in a timescale $1/\gamma$ [e.g. the first two emission events in the individual quantum trajectory depicted in Fig.~\ref{fig:fig3-heralding}(a)].
We can improve the protocol to work for an arbitrary $h$ by applying a heralding pulse implementing the rotation $\hat U=\exp(-i\theta\hat\sigma_y)$ just after the activation pulse, which transforms the emitter as $|\pm\rangle \to |g/e\rangle$ and enables the optical decay conditioned on the prior emission of a THz photon.

The protocol can establish a near-perfect correlation between the number of photons emitted in the THz and visible regimes, as shown in Fig.~\ref{fig:fig3-heralding}(b,c). This correlation enables accurate reconstruction of the photon number distribution $p(n)$ of the emitted THz state directly from optical measurements of its visible counterpart.
If the photon-number resolution  necessary to reconstruct $p(n)$ in the visible is not accessible, the protocol still allows one to herald a single THz photon emitted during the activation pulse interval $t\in(0,\tau)$, by the detection of an optical photon emitted after the pulse, i.e., at $t>\tau$. To quantify the efficiency of this heralded emission process, we define the corresponding  conditional probability $E \equiv P(\text{THz}_{<\tau}|\text{optical}_{>\tau})$, (details on the calculation are provided in the SM). The result is shown as blue circles in Fig.~\ref{fig:fig2}(b,d), indicating that this heralding protocol yields high  efficiency $E\sim 0.92-0.95$, provided that  the activation pulse duration is much longer than the cavity-photon lifetime $\tau \gg \kappa^{-1}$ and does not exceed $\tau_\text{max}$. This latter upper limit arises since this simpler heralding does not discriminate between single- and multi-THz-photon emission, and therefore the heralding efficiency is degraded by optical repumping for $\tau > \tau_\text{max}$ in the same way it degrades $P_1(\tau)$. 

Optical heralding allows the use of shorter activation pulses to improve purity at the expense of brightness, 
 enabling near-perfect heralded single-photon emission even when the intrinsic limit $P_\text{max}$ is constrained. See Fig.~\ref{fig:fig2}(d), where $\text{max}(E)\approx 0.95 \gg P_\text{max}$ and $I$ reaches 91\%.
 This technique provides no advantage for red detuning ($\Delta<0$), where the heralding efficiency is limited by the single-photon probability, i.e. $E(\tau)\approx P_1(\tau)$ for $\tau\gg(\gamma+\Gamma)^{-1}$.

\emph{Experimental feasibility---}
Unlike ensemble-based lasing setups~\cite{chestnov2017a}, which benefit from a collective enhancement of the light--matter interaction, the present proposal relies entirely on optimizing the single-emitter cooperativity $\tilde C$ through a careful choice of cavity parameters (influencing $\chi$ and $\kappa$), permanent dipole moment (influencing $\chi$), and optical transition dipole moment (influencing $\gamma$).  For the permanent dipole moment, we have taken a reference value of $|\mathbf{d}_{ee}|= 250\,\text{D} \approx e \cdot 5\,\text{nm}$, close to the values of a few $e\cdot \text{nm}$ that have been reported in nitride-based quantum dots~\cite{ostapenko2010, kindel2014}. It is also consistent with the electron-hole separation achieved in colloidal semiconductor quantum dot molecules~\cite{ossia2023}, and below the large permanent dipole moment achieved in self-assembled quantum dot molecules, which can reach electron-hole separations of tens of nm~\cite{muller2012,wang2009}. Large permanent dipole moments offer an added benefit to source performance by exhibiting longer radiative lifetimes due to the reduced overlap between electron and hole wavefunctions~\cite{lv2021a}.

This value of $|\mathbf{d}_{ee}|$ leads to  coupling rates to the cavity modes $L_1$ and $L_3$ in the GHz range (see Table~\ref{tab:parameters}). We set the Rabi frequency  to $\Omega/2\pi=0.2$ THz, in line with previous experimental reports with semicondutor quantum dots~\cite{kim2014,boos2024}. This choice requires different detunings $|\Delta|$ in each configuration to satisfy the resonance condition $\Omega_R = \omega_c$, yielding dressing ratios of $h = 0.0325$  for $L_1$ and $h = 0.01955$ for $L_3$. Finally, we set $\gamma/2\pi = 39.7\,\text{MHz}$, giving a lifetime of 4 ns typical for solid-state quantum emitters~\cite{esmann2024}.
These parameters result in effective cooperativities $ \tilde{C} = 1.94 $ (for $ L_1$), and $ \tilde{C} = 24.65 $ (for $ L_3$), yielding a $P_\text{max}$ of $\sim 60\%$ and 90\%, respectively.

Regarding sources of decoherence beyond spontaneous emission, we have shown (see Sec.~IX of the SM) that thermal noise and pure dephasing should have a negligible effect under standard helium cooling conditions. 

Finally, the single-photon sensitivity required to directly verify our THz SPE in a HBT setup has been demonstrated in several classes of detectors, 
such as quantum dot SET detectors ($\sim$0.5--3 THz)~\cite{komiyama2000}, charge-sensitive infrared phototransistors ($\sim$6--30 THz)~\cite{an2007}, and quantum-capacitance detectors (QCD) ($\sim$1.5 THz)~\cite{echternach2018}.

\emph{Conclusions---} We have proposed an on-demand THz single-photon source featuring wide frequency tunability and optical heralding with efficiencies between 90-95\% comparable to state-of-the-art nondeterministic parametric down conversion sources \cite{tinsley2016,kaneda2019,Leontyev2021,bock2016}. Using our custom-designed hybrid THz cavities combined with optical quantum emitters with experimentally available permanent dipole moments, the scheme achieves high efficiencies (65-92\%) and purities  (88-100\%) comparable to the best visible single-photon sources \cite{esmann2024}, requiring only standard helium cooling and optical lasers. The mechanism proposed enables heralding and complete characterization of photon statistics of the emitted THz state via detection of photons in the visible regime. 
Our source demonstrates the potential of optically dressed polar quantum emitters to enable the first experimental realizations of deterministic THz single-photon sources and to serve as a building block for developing quantum optics in this frequency regime.\\[10pt]

\begin{acknowledgments}

This work makes use of the Quantum Toolbox in Python
(QuTiP)~\cite{johansson2012,johansson2013}. We acknowledge financial
support from the Proyecto Sin\'ergico CAM 2020 Y2020/TCS-
6545 (NanoQuCo-CM), and MCINN projects PID2021-126964OB-I00 (QENIGMA), TED2021-130552B-C21 (ADIQUNANO) and PID2024-156077OB-I00 (DQUOTE).
C. S. M. acknowledges support by the project PID2023-149969NA-100 funded by the Spanish Agencia Estatal de Investigación MICIU/AEI/10.13039/501100011033, and by a 2025 Leonardo Grant for Scientific Research and Cultural Creation from the BBVA Foundation. C. S. M. and D. M. C. also acknowledge 
the support of a fellowship from la Caixa Foundation (ID 100010434), from the European Union's Horizon 2020 Research and Innovation Programme under the Marie Sklodowska-Curie Grant Agreement No. 847648, with fellowship codes  LCF/BQ/PI20/11760026 and LCF/BQ/PI20/11760018. D. M. C. also acknowledges support from the Ramon y Cajal program (RYC2020-029730-I). C. G. acknowledges support from the Knut and Alice Wallenberg Foundation through the Wallenberg Centre for Quantum Technology (WACQT). All authors thank Antonio I. Fernández-Domínguez for  fruitful discussions and initial simulation support in the project.
\end{acknowledgments} 

\bibliography{DeterministicSingleTHzphoton, Permanent-dipole-QDs, Detection}

\clearpage
\onecolumngrid
\begin{center}
{\bf \large Supplementary Material}
\end{center}

\renewcommand{\theequation}{S\arabic{equation}}

\renewcommand{\thefigure}{S\arabic{figure}} 
\setcounter{figure}{0} 
\setcounter{equation}{0}   

\label{AppendixA}

\setcounter{secnumdepth}{2}


\onecolumngrid
\section{Indistinguishability and purity}
\label{indistinguishability}

One measure of the quality of a single-photon source is the indistinguishability of the emitted photons. The indistinguishability  can be verified by a Hong-Ou-Mandel setup, where two subsequent one-photon pulses from the source are made to interfere on a beam splitter. The crosscorrelations of the two output channels are~\cite{hughes2019}:
\begin{equation}
    G_\text{HOM}(t,t')=\frac{1}{2}(G^{(2)}_\text{pop}(t,t')+G^{(2)}(t,t')-|G^{(1)}(t,t')|^2),
\end{equation}
where the first- and second-order correlation functions employed are defined as
\begin{eqnarray}
        G^{(2)}_\text{pop}(t,t')&\equiv &\langle\hat a^\dagger\hat a\rangle(t+t')\langle\hat a^\dagger\hat a\rangle(t),\\
        G^{(1)}(t,t')&\equiv &\langle\hat a^\dagger(t)\hat a(t+t')\rangle,\\
        G^{(2)}(t,t')&\equiv &\langle\hat a^\dagger(t)\hat a^\dagger(t+t')\hat a(t+t')\hat a(t)\rangle.
\end{eqnarray}
The indistinguishability (or two-photon interference visibility) can then be computed by taking the ratio of the interference peaks at $t'=0$ and $t'=2\tau$ of a pulsed source with period $2\tau$~\cite{hughes2019}:
\begin{equation}
        I=1-\frac{\int_0^{\tau}dt\int_{-{\tau}}^{\tau}dt' G^{(2)}_\text{HOM}(t,t')}{\int_0^{\tau}dt\int_{\tau}^{3{\tau}}dt' G^{(2)}_\text{HOM}(t,t')}=1 -\frac{\int_0^{\tau}dt\int_0^{\tau}dt'(G^{(2)}_\text{pop}(t,t')+G^{(2)}(t,t')-|G^{(1)}(t,t')|^2)}{\int_0^{\tau}dt\int_0^{\tau}dt'(2G^{(2)}_\text{pop}(t,t')-|\langle\hat a^\dagger(t)\rangle\langle\hat a(t+t')\rangle|^2)}.
\end{equation}
This definition is the one adopted in the results for $I$ shown in this text, which we compute via the quantum regression theorem. To obtain the purity, we compute the second order degree of coherence over the pulse by dividing the two photon probability by the squared integrated photon probability (both integrated over the pulse time), i.e.

\begin{equation}
    g^{(2)}(0)=\frac{\int_0^{\tau}dt\int_{0}^{\tau}dt' G^{(2)}(t,t')}{\int_0^{\tau}dt\int_{0}^{\tau}dt' G^{(2)}_\text{pop}(t,t')}.
\end{equation}
In a regime where we are close to single-photon generation the classical coherent amplitude should vanish, i.e., $|\langle\hat a^\dagger(t)\rangle\langle\hat a(t+t')\rangle|^2\approx0$. Using our adiabatic model $\mathcal{L}_\text{adb}$, we can evaluate the correlation functions to be
\begin{eqnarray}
        G^{(2)}_\text{pop}(t,t')&\sim &\langle\hat \zeta_+\hat \zeta_-\rangle(t+t')\langle\hat \zeta_+\hat \zeta_-\rangle(t)=\begin{cases}
			\frac{ (1+\tilde C e^{-\gamma(1+\tilde C)t})(1+\tilde C e^{-\gamma(1+\tilde C)(t+t')})}{(1+\tilde C)^2}, & \Delta>0\\
            e^{-\gamma(1+\tilde C)(2t+t')}, & \Delta<0
		 \end{cases},\\
        G^{(1)}(t,t')&\sim &\langle\hat \zeta_+(t)\hat \zeta_-(t+t')\rangle=\begin{cases}
			\frac{  e^{-\gamma(1+\tilde C)t'/2}(1+\tilde C e^{-\gamma(1+\tilde C)t})}{1+\tilde C}, & \Delta>0\\
            e^{-\gamma(1+\tilde C)(2t+t')/2}, & \Delta<0
		 \end{cases},\\
        G^{(2)}(t,t')&\sim &\langle\hat \zeta_+(t)\hat \zeta_+(t+t')\hat \zeta_-(t+t')\hat \zeta_-(t)\rangle=\begin{cases}
			\frac{ (1- e^{-\gamma(1+\tilde C)t'})(1+\tilde C e^{-\gamma(1+\tilde C)t})}{(1+\tilde C)^2}, & \Delta>0\\
            0, & \Delta<0
		 \end{cases}.
\end{eqnarray}
From this, one can see that that $G^{(2)}(t,t')\approx G^{(2)}_\text{pop}(t,t')-|G^{(1)}(t,t')|^2$, which leads to the purity following the indistinguishability.

\section{Analytical Expressions for the Maximum of the Single-Photon Probability}
\label{analytical}

\subsection{Blue detuning $\Delta>0$}

We can derive an explicit expression for $P_1(\tau)$ using the simplified model given by $\mathcal{L}_\text{adb}$ ($\mathcal{K}_\text{adb}=\Gamma\hat\zeta_-\hat\rho\hat\zeta_+$), i.e.
\begin{equation}
\begin{split}
P_1(\tau)
     &=\int_0^\tau\text{Tr}[e^{(\mathcal{L}_\text{adb}-\mathcal{K}_\text{adb})(\tau-t)}\mathcal{K}_\text{adb}e^{(\mathcal{L}_\text{adb}-\mathcal{K}_\text{adb})t}\hat\rho_0] dt=\int_0^\tau\text{Tr}\left[e^{(\mathcal{L}_\text{adb}-\mathcal{K}_\text{adb})(\tau-t)}\mathcal{K}_\text{adb}\begin{pmatrix}
             e^{-\gamma\tilde C t}&0\\
             0&0
         \end{pmatrix}\right] dt\\
         &=\int_0^\tau\text{Tr}[e^{(\mathcal{L}_\text{adb}-\mathcal{K}_\text{adb})(\tau-t)}\begin{pmatrix}
             0&0\\
             0&\gamma\tilde C e^{-\gamma\tilde C t}
         \end{pmatrix}] dt=\int_0^\tau \text{Tr}\begin{pmatrix}
             -\frac{\gamma^2\tilde Ce^{-t\gamma \tilde C}(e^{-(\tau-t)\gamma }-e^{-(\tau-t)\gamma \tilde C})}{\gamma-\gamma\tilde C}&0\\
             0&\gamma\tilde Ce^{-t\gamma \tilde C}e^{-(\tau-t)\gamma }
         \end{pmatrix}dt\\
         &=\frac{e^{-\tau\gamma}-e^{-\tau\gamma \tilde C}[1+\gamma \tau(1-\frac{1}{\tilde C})]}{(1-\frac{1}{\tilde C})^2}.
         \label{Panalytics}
     \end{split}
\end{equation}

\noindent The exact solutions to the time and amplitude of the maximum of the single-photon emission probability in Eq.~(1)
are obtained by rewriting the condition $dP_1(\tau)/d\tau=0$ as
\begin{equation}
    \gamma\tau_\text{max}=-1-\frac{1}{\tilde C(\tilde C-1)}+\frac{1}{\tilde C-1}e^{\gamma\tau_\text{max}(\tilde C-1)},
\end{equation}
which is of the form $x=\alpha+\beta e^{\delta x}$ and has the general solution $x=\alpha-W(-\beta\delta e^{\alpha\delta})/\delta$, where $W$ is the product logarithm (Lambert $W$ function)~\cite{corless1996}. 
This yields
\begin{equation}
     \tau_\text{max}=\frac{\gamma^{-1}}{1-\tilde C}\left(
W_{-\theta(\tilde C)}(\Xi) + \frac{1}{\tilde C}+\tilde C -1
    \right)
\end{equation}
and
\begin{equation}
P_1(\tau_\text{max})=\exp\left[{\frac{\frac{1}{\tilde C}+W_{-\theta(\tilde C)}(\Xi)}{\tilde C-1}}\right]\frac{\tilde C+W^{-1}_{-\theta(\tilde C)}(\Xi)}{e^{-1}(\tilde C-1)},
\end{equation}
where $\Xi=-e^{1-\tilde C-\frac{1}{\tilde C}}$ and $\theta(x)$ is the Heaviside step function. $W_0(x)$ and $W_{-1}(x)$ are the two real branches of the product logarithm.
 Evaluating $W_r(\Xi)$ for large or small values $\tilde C$ can be problematic, since in the limiting case $\Xi\rightarrow 0$ the functions $W_{-1}(0)$ and $1/W_0(0)$ are not defined. In the following, we present some useful approximations that can be used in these limits.
\subsection{Limit $\tilde C\gg1$}
For large values of $\tilde C$ we can use the following upper bound as an approximation $W_{-1}(x)\leq\ln(-x)-\ln[-\ln(-x)]$~\cite{corless1996}. This yields a simplified expression $\gamma \tau_\text{max}\approx\ln\tilde C/\tilde C$, which we can rewrite as
\begin{equation}
  \tau_\text{max}\approx \Gamma^{-1}\ln\tilde C
    \end{equation}
and similarly for the maximum single-photon probability:
\begin{equation}
P_1(\tau_\text{max})\approx 1-\frac{\ln\tilde C}{\tilde C}.
\end{equation}
\subsection{Limit $\tilde C\ll1$}
For small values of $\tilde C$, we use a first-order Taylor expansion around 0, $W_0(x)=x+\mathcal{O}(x^2)$~\cite{corless1996}. This yields:
\begin{equation}
    \tau_\text{max}\approx \Gamma^{-1}
\end{equation}
and
\begin{equation}
P_1(\tau_\text{max})\approx e^{-1},
\end{equation}
where we used that $ -\Xi\ll 1/\tilde C$.

\subsection{Limit $\tilde C=1$}
Although $\tau_\text{max}$, $P_1$ and $P_1(\tau_\text{max})$ are undefined in $\tilde C=1$, the limits are well-defined, yielding
\begin{eqnarray}
\lim_{\tilde C\rightarrow1} P_1(\tau)&=&\gamma\tau(1+\gamma\tau/2)e^{-\tau\gamma},\\       
\lim_{\tilde C\rightarrow1}\tau_\text{max}&=&\sqrt{2}/\gamma,\\
\lim_{\tilde C\rightarrow1}P_1(\tau_\text{max})&=&(1+\sqrt{2})e^{-\sqrt{2}}.   
\end{eqnarray}

\subsection{Limit $\gamma\gg\kappa,g$}
The analytics in this limit remain mostly the same except that the roles of $\kappa$ and $\gamma$ are switched, as can be seen by adiabatically eliminating the emitter, i.e. $\mathcal{L}_\text{adb}{\hat\rho}=\frac{\kappa}{2}\mathcal{D}(\hat a)\hat\rho+\frac{2g^2}{\gamma}\mathcal{D}(\hat a^\dagger)\hat\rho$. This yields
\begin{eqnarray}
         P(\tau)&=&\frac{e^{-\tau\kappa}-e^{-\tau\kappa \tilde C}[1+\kappa \tau(1-\frac{1}{\tilde C})]}{(1-\frac{1}{\tilde C})^2},\\
         P_0(\tau)&=&\exp(-\kappa\tilde C\tau),\\
         \kappa\tau_\text{max}&=&-1-\frac{1}{\tilde C(\tilde C-1)}-\frac{1}{\tilde C-1}W_{-\theta(\tilde C)}(\Xi).
\end{eqnarray}

\subsection{Red detuning $\Delta<0$}
 We can compute the zero-photon probability as before
\begin{equation}
    P_0(\tau)=\int_0^\tau\text{Tr}(S_\tau\hat\rho_0) dt=\frac{1}{\tilde C+1}(1+\tilde Ce^{-\tau(\gamma+\Gamma )}),
\end{equation}
from which we can obtain the complementary single-photon probability since the lack of the previously present incoherent pumping process removes the chance of repopulating the $|+\rangle\langle+|$ state and emission of a second THz photon
\begin{equation}
    P_1(\tau)=1-P_0(\tau)=\frac{\tilde C}{\tilde C+1}[1-e^{-\tau(\gamma+\Gamma )}].
\end{equation}
$P_1$ is maximized for large times $\tau$, i.e., there is no longer a timing issue. The maximal value is given by
\begin{equation}
    P_\text{max}=\frac{\tilde C}{\tilde C+1}.
\end{equation}

\section{The near-resonant regime}
\label{sec:appendix-near-resonant}
The near-resonant regime yields larger coupling strengths, which may suggest larger single-photon probabilities and shorter emission timescales. However, the subsequent change in the dressing ratio compromises the assumption that $\gamma_\pm$ can be neglected, and the effective cooperativity becomes more accurately described \cite{groiseau2024} by
\begin{equation}
    \tilde C=\frac{4g^2}{\kappa(\gamma_-+\gamma_+)}=\frac{4C}{h^2+h^{-2}}.
\end{equation}
Assuming that the adiabatic elimination still holds, the general single-photon probability is  given by
%

\begin{equation}
      P_1(\tau) = \frac{-2g^2}{A}  \frac{e^{-\tau (\gamma_-+\gamma_++\frac{4g^2}{\kappa})}}{[4g^2 + (\gamma_- + \gamma_+)\kappa]^2 -16g^2\gamma_+\kappa } \left[f(A) - f(-A) \right],
      \label{nearresonantanalytics}
\end{equation}
where
\begin{equation}
    A \equiv \sqrt{16g^4 + 8g^2(\gamma_- - \gamma_+)\kappa + (\gamma_- + \gamma_+)^2 \kappa^2},
\end{equation}
and
\begin{multline}
    f(A)\equiv e^{\frac{\tau}{2\kappa}(4g^2 + \gamma_- \kappa + \gamma_+ \kappa - A)}\\
      \times\bigg[\kappa (\gamma_- + \gamma_+) \left[ \gamma_+ \tau \left( \kappa (\gamma_- + \gamma_+) - A \right) + 2 \gamma_- \kappa \right] 
\\+ 4g^2 \left[ \gamma_+ \tau \left( A + 2 \gamma_- \kappa - 2 \gamma_+ \kappa \right) + 4 \gamma_- \kappa - 2 \gamma_+ \kappa \right] 
+ 16g^4 (\gamma_+ \tau + 2) \bigg].
\end{multline}

\begin{figure}[H]
\centering
	\includegraphics[width=0.7\linewidth]{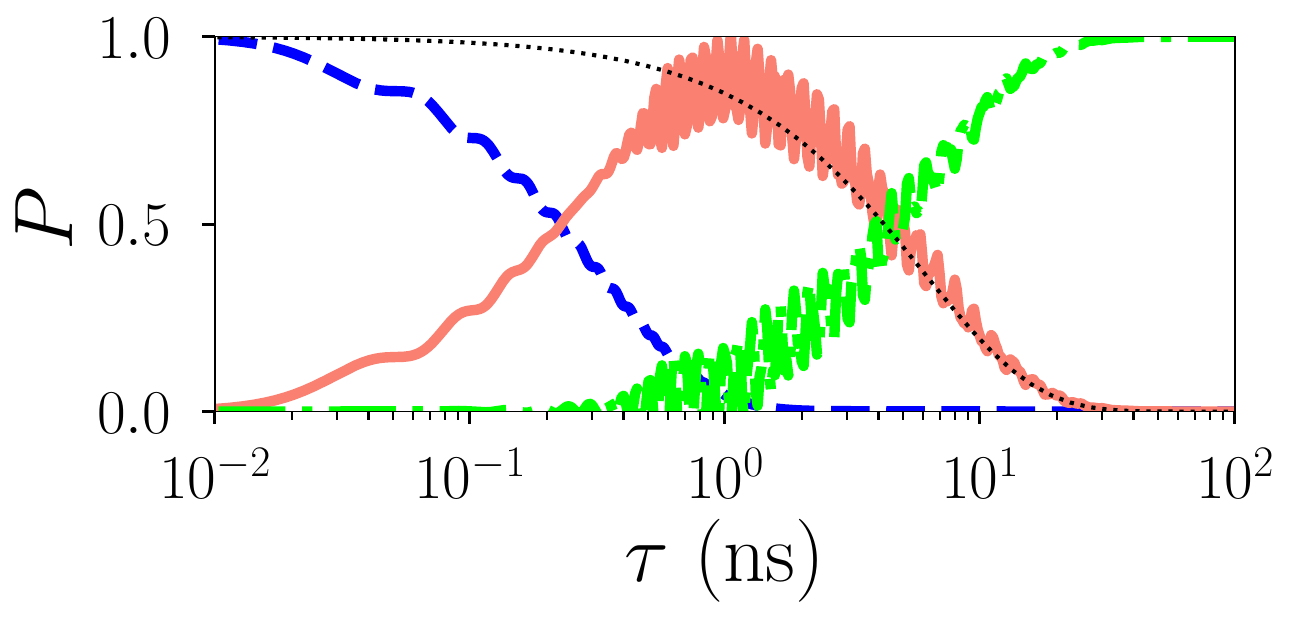}
     \caption{Single-photon probability for a near-resonant drive $\Omega \sim \Omega_R$. 
     The numerically computed single-photon probability $P_1(\tau)$ (solid red) is shown alongside the analytical prediction (dotted black) and the numerical probabilities of detecting zero photons $P_0(\tau)$ (dashed blue) and more than one photon $P_{>1}(\tau)$ (dash-dotted green), for the parameter set  $\chi/2\pi = 12.5\,\text{GHz}$, 
$\kappa/2\pi = 0.974\,\text{GHz}$, 
$\gamma/2\pi = 39.78\,\text{MHz}$, 
$\Omega/2\pi = 4\,\text{THz}$, 
$\Delta/2\pi = 3.19\,\text{THz}$, and 
$\omega_c/2\pi = 5.116\,\text{THz}$.} 
    \label{example2}
\end{figure}

In the resonant limit, we have $\gamma_-\approx\gamma_+\approx\gamma/4$ and $g\approx\chi$ ($\tilde C=2C$), which we can use to simplify further the expression of the single-photon probability,
\begin{multline}
        P_1(\tau)=\frac{e^{-\gamma\tau(1+\frac{C}{2}+\sqrt{1+4C^2})}}{(1+4C^2)^\frac{3}{2}}
        \bigg\{C^3(-1+e^{\gamma\tau\sqrt{1+4C^2}})(8+\gamma\tau)
        -\frac{C^2}{2}\left[4(-1+e^{\gamma\tau\sqrt{1+4C^2}})
        - \right.\\
        \left.(1+e^{\gamma\tau\sqrt{1+4C^2}})\gamma\tau\sqrt{1+4C^2}\right]
        +\frac{C}{4}\left[(-1+e^{\gamma\tau\sqrt{1+4C^2}})(4+\gamma\tau) \right.\\
        \left.
        +(1+e^{\gamma\tau\sqrt{1+4C^2}})\gamma\tau\sqrt{1+4C^2}\right]\bigg\}.
\end{multline}
Figure~\ref{example2} shows numeric and analytic results for the single-photon probability close to resonance. Since $g/\kappa$ ratio is larger than in the regime of parameters considered in the main text, the analytical results are worse at capturing the fast transient rise, failing to capture the oscillations of $P_1$ in time due to vacuum Rabi oscillations of the THz photon. Nevertheless, equation~(\ref{nearresonantanalytics}) still matches the average $P_1$ pretty well in the long time limit. The oscillations introduce fluctuations in the optimal pulse length $\tau_{\text{max}}$, accounting for its non-monotonic behavior near resonance in Fig.~3.

\section{The off-resonant regime}
\label{sec:appendix-off-resonant-regime}
\textcolor{blue}{For $\Delta>0$,} operating in the off-resonant regime offers several advantages that may compensate for the reduction in effective coupling strength. One clear benefit is that achieving generalized Rabi frequencies $\Omega_R$ well within the THz range becomes feasible with lower driving amplitudes $\Omega$, and therefore reduced optical power. Another advantage is that the required initial state becomes close to the ground state of the bare qubit, since for $\Delta \gg \Omega$, we find $|\langle g|+\rangle|=c\approx1$. This means that, for a detuning $\Delta$ large enough, one could simplify the experimental protocol by omitting the first preparation pulse that initializes the state $|+\rangle$. Figure~3 shows the performance of the scheme when dispensing the preparation pulse and when using $|g\rangle$ as the initial state. A deviation from the results obtained with the optimal initial state $|+\rangle$ is only noticeable for $h>10^{-1}$, i.e. $1-\Delta/\Omega_R>10^{-2}$, which corresponds to $c\approx1+(\Delta/\Omega_R-1)/4\lesssim0.998$. Assuming that we can achieve a $C$ large enough that $\tilde C\gg10$, we can solve $P_\text{max}=1-\ln\tilde C/\tilde C$ for $\tilde C$ and obtain
\begin{equation}
    \tilde C=\exp[-W_{-1}(P_\text{max}-1)].
\end{equation}
Using this equation and that outside of the near-resonant regime, we can approximate $\tilde C\approx 4s^2C\approx C\Omega^2/\Omega_R^2 \approx C\Omega^2/\Delta^2$, we can find the bounds for, e.g. $C$ or $\Delta$ that could achieve a certain $P_\text{max}$ depending on what we assume to be fixed.

\begin{figure}[t]
    \includegraphics[width=0.7\linewidth]{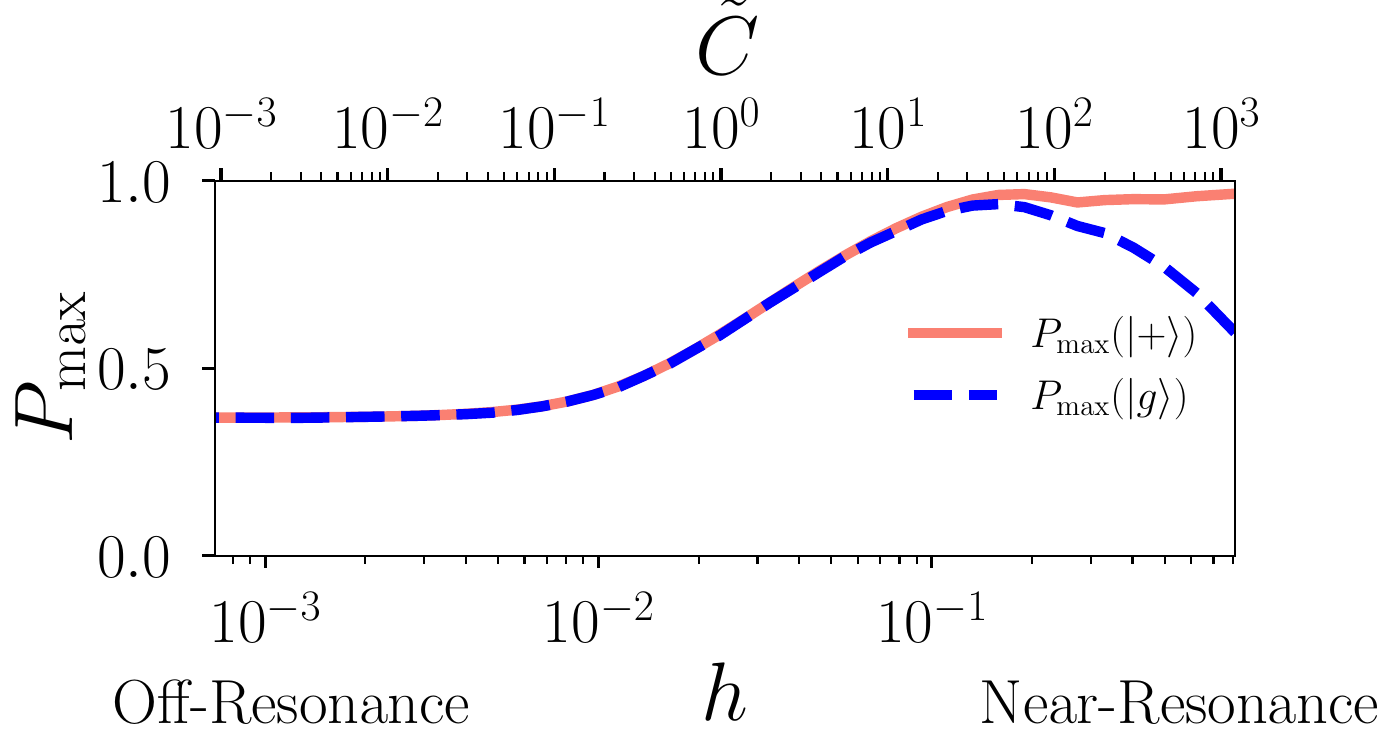}
    \caption{\textbf{Single-photon probability for different initial states}. Maximum single-photon detection probability $P_\text{max}$, obtained numerically (solid red) as a function of the detuning $\Delta$ for the parameter set $\{\chi,\kappa,\gamma,\Omega,\Omega_R\}/2\pi=\{4,3.5,39.78,0.2,3078\}$ GHz. The dashed blue line in the bottom plot represents the situation where the initial state is $|g\rangle$.} 
\end{figure}

\section{Efficiency vs. Purity}
\begin{figure}[h]
        \includegraphics[width=0.7\linewidth]{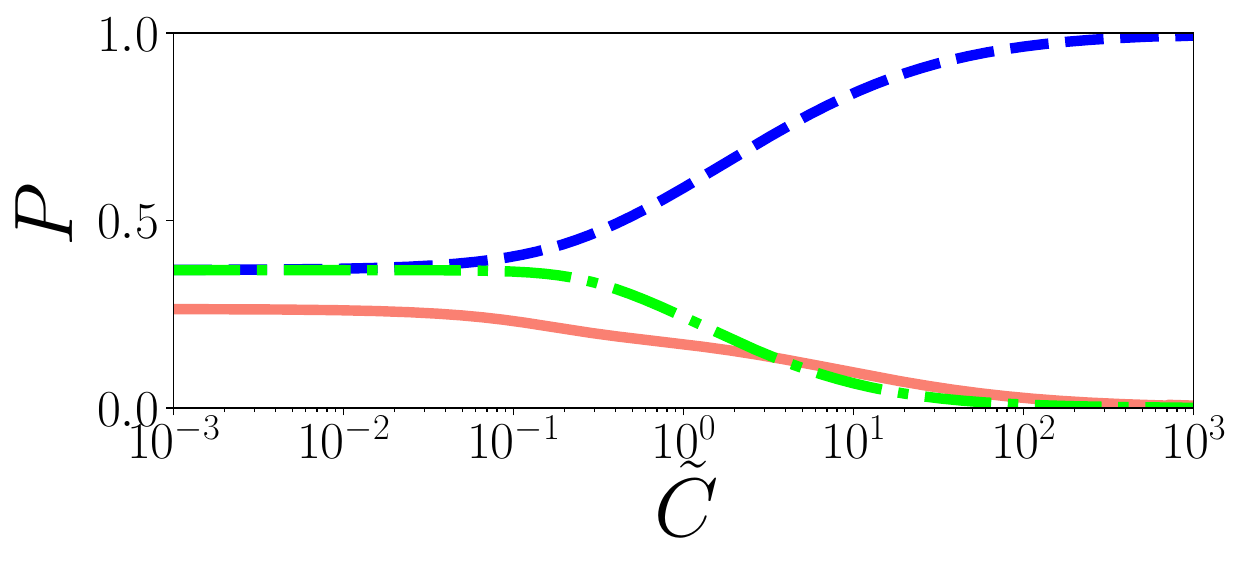}
    \caption{Plot of the maximum single-photon detection probability $P_1(\tau_\text{max})$ (dashed blue), $P_0(\tau_\text{max})$ (dashdotted green) and $P_{>1}(\tau_\text{max})$ (solid red) as a function of $\tilde C$.} 
    \label{P0}
\end{figure}

For $\Delta>0$, whether the main limitation for perfect single-photon generation is efficiency or purity depends on the effective cooperativity $\tilde C$. Fig.~\ref{P0} shows the dependence of the probabilities of measuring zero [$P_0(\tau_\text{max})$], one [$P_\text{max}=P_1(\tau_\text{max})$], or more than one [$P_{>1}(\tau_\text{max})$] photons at the optimal pulse length $\tau_\text{max}$ as a function of the effective cooperativity $\tilde C$.

We find that $P_{>1}(\tau_\text{max})>P_0(\tau_\text{max})$ in the limit of large $\tilde C$, while the opposite is true for small $\tilde C$. In the relevant limit of large $\tilde C$,
where $P_\text{max} \to 1-\ln \tilde C/\tilde C$, the vacuum and multiphoton sources of imperfection are given by
\begin{equation}
    P_0(\tau_\text{max}) = \frac{1}{\tilde{C}}, \quad
P_{>1}(\tau_\text{max}) =  \frac{\ln \tilde{C} - 1}{\tilde{C}}.
\end{equation}
 Assuming that at $\tau_\text{max}$ the two-photon probability $P_2(\tau_\text{max})$ is roughly equal to $P_{>1}(\tau_\text{max})$, and that detecting one photon is by far the most likely outcome $P\gg P_0(\tau_\text{max}),P_{>1}(\tau_\text{max})$, the single-photon purity can be estimated as 
 \begin{equation}
     \Pi(\tau_\text{max})=1-g^{(2)}(0)\approx 1-2\frac{\ln\tilde C}{\tilde C}.
 \end{equation}
 In the limit of low $\tilde C$, we find $P_{>1}\rightarrow1-2e^{-1}$.

\section{Realistic Laser Pulse Shapes}
We assume throughout this work that the various pulses are turned on and off instantaneously, which is a justified approximation since the typical rise and fall times are of the order of hundreds of ps and hundreds of fs \cite{sussman2006,goban2008}, respectively. This time is much shorter than the expected decay in our setups, which are of the order of ns or more.
Nevertheless, given that experimental laser pulses do not exhibit the idealized square form assumed in this work, in this section we investigate the influence of a more realistic time-dependent pulse that features a finite rise and fall time given by $\tau_p$, and that is turned off at a time $\tau$:
\begin{equation}
    \Omega(t)=\begin{cases}
        1-e^{-\frac{t}{\tau_p}}& t<\tau\\
        e^{-\frac{t}{\tau_p}}& t>\tau.
    \end{cases}
\end{equation}
Importantly, this time-dependence affects the dynamical composition of the dressed states $|\pm\rangle$, the coupling $g$, and the spontaneous processes $\gamma_{\pm,z}$. We base our simulations on quantum trajectories of the effective Jaynes-Cummings model with $\hat H_{\text{JC}}$.

\begin{figure}[h]
        \includegraphics[width=1\linewidth]{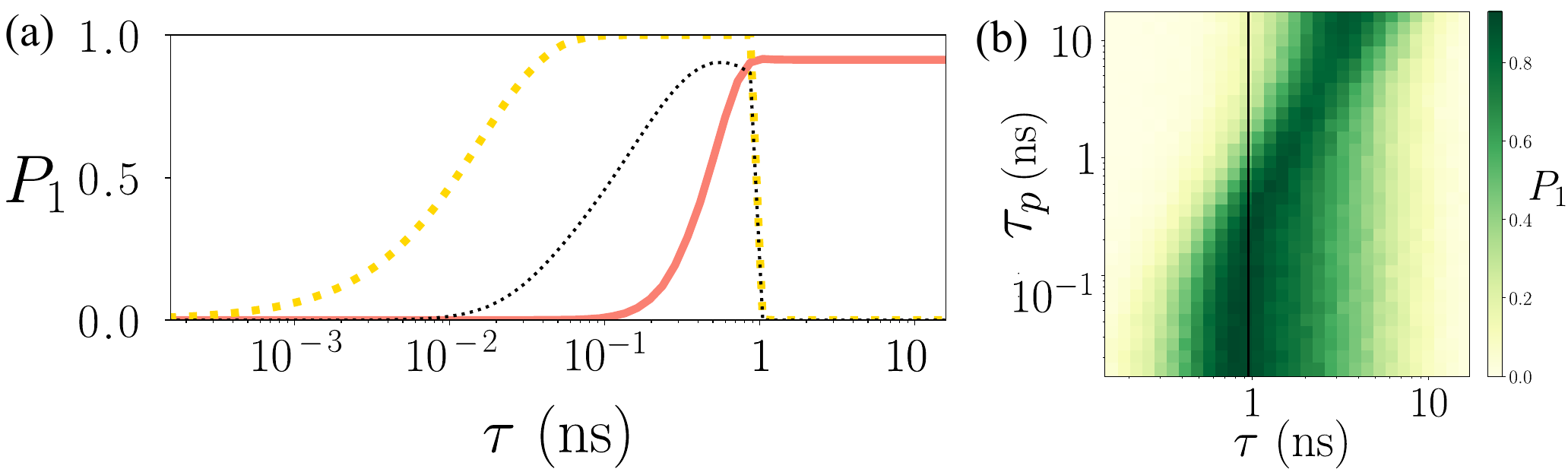}
    \caption{\textbf{Single-photon probability for non-square pulses}. (a) $P_1(\tau)$ obtained numerically from $10^4$ quantum trajectories (solid red) and analytically (dotted black) and the temporal profile of the laser drive (dotted yellow) for the parameter set $\{\chi,\kappa,\gamma,\Omega,\Omega_R,\tau_p^{-1}\}/2\pi=\{12.5,0.974,39.79\times10^{-3},0.2,5116,10^{-2}\}$ GHz. The shut off is not instantaneous (even though it might appear so in the plot due to the logarithmic scale). (b) Map of the Maximum single-photon detection probability obtained numerically from 1000 quantum trajectories, as a function of the pulse length $\tau$ and the raising time $\tau_p$ for the parameter set $\{\chi,\kappa,\gamma,\Omega,\Omega_R\}/2\pi=\{12.5,0.974,39.79\times10^{-3},0.2,5116\}$ GHz. The black line represents $\tau_\text{max}$ for instantaneous rise and fall times. } 
    \label{timedependent}
\end{figure}

Figure~\ref{timedependent}(a) shows that a finite rise time delays the onset of $P_1(\tau)$ for $\tau<\tau_\text{max}$,  compared to the analytical results assuming an instantaneous rise. In addition, Fig.~\ref{timedependent}(b) shows the map $P_1(\tau)$ as a function of the pulse length and the rise time. We conclude from this map that the timescale of the rise time of the laser pulse has no noticeable influence on the dynamics of the single-photon probability as long as $\tau_\text{max}>\tau_p$. However, once both timescales are comparable, we observe a temporal shift of the maximum $P_1$ to larger times. The dynamics at large $\tau$  are not affected, as the decay rates are typically much lower than the inverse fall time $\gamma,\Gamma\ll\tau_p^{-1}$. Typical rise times are of the order of hundreds of ps, while typical fall times are of the order of hundreds of fs \cite{sussman2006,goban2008}. Since the expected decay time in realistic setups will be of the order of at least ns, we conclude that considering the laser ramp as effectively instantaneous is justified.

\section{Heralding}
While spontaneous emission at optical frequencies tends to degrade single-photon emission
due to repumping of the quantum emitter and subsequent multi-photon contributions, it also provides a potential pathway for optical heralding of THz photon emission, as discussed in detail in this section.

In every individual realization of the experiment (a single trajectory in the quantum trajectory picture), a system that has not undergone optical spontaneous emission (no optical photodetection) is in one of these two possible states:
\begin{itemize}
    \item Before THz-photon emission from the THz cavity: 
    \begin{equation*}
        |\psi_1\rangle = \alpha|+\rangle|0\rangle_\text{THz}+\beta|-\rangle|1\rangle_\text{THz},\quad (\alpha\gg\beta)
    \end{equation*}
    \item After THz-photon emission from the THz cavity:
    \begin{equation*}
        |\psi_2\rangle = |-\rangle|0\rangle_\text{THz}.
    \end{equation*}
\end{itemize}
We recall that the emission of a THz photon is given by the collapse superoperator
\begin{equation}
    \mathcal K \hat\rho \equiv \kappa \hat X \hat \rho \hat X^\dagger.
\end{equation}
On the other hand, spontaneous emission of an optical photon is described by the superoperator 
\begin{equation}
     \mathcal K_\text{opt}\hat\rho\equiv\gamma\hat\sigma_-\hat\rho\hat\sigma_+
\end{equation}
which is the only way the system can go from $|-\rangle$ to $|+\rangle$, thus restarting the cycle and allowing to generate more THz photons. Therefore checking that no optical photon is emitted before turning off the laser at a time $\tau$ guarantees the emission of at most one single THz photon. 

In order to determine whether a THz photon has been emitted in the first place, we can apply a second pulse after turning off the laser at time $\tau$. The pulse must implement the unitary transformation 
\begin{equation}
    \hat U=\exp(-i\theta\hat\sigma_y)
\end{equation}
which transforms the emitter as
$|\pm\rangle \to |g/e\rangle$, and therefore the states as
\begin{eqnarray}
    |\psi_1\rangle &\to& \alpha|g,0\rangle_\text{THz}+\beta|e,1\rangle_\text{THz},\\
        |\psi_2\rangle &\to& |e,0\rangle_\text{THz}.
\end{eqnarray}
We can write the action of this pulse in superoperator form as 
\begin{equation}
    \mathcal U \equiv \hat U \hat\rho \hat U^\dagger.
\end{equation}

After this pulse, we let the system evolve freely described under the Liouvillian $\mathcal{L}_0\hat\rho=\frac{\kappa}{2}\mathcal{D}[\hat a]\hat\rho+\frac{\gamma}{2}\mathcal{D}[\hat \sigma_-]\hat\rho$ for a time long enough to observe all potential decays, e.g. $\tau'=30/\gamma$. It is straightforward to see that the emission of an optical photon due to the relaxation from the excited state $|e\rangle$ is only compatible with states in which a THz photon was put into the cavity (which is eventually emitted), thus heralding the emission of the THz photon.

\begin{figure}[b]
\includegraphics[width=0.95\linewidth]{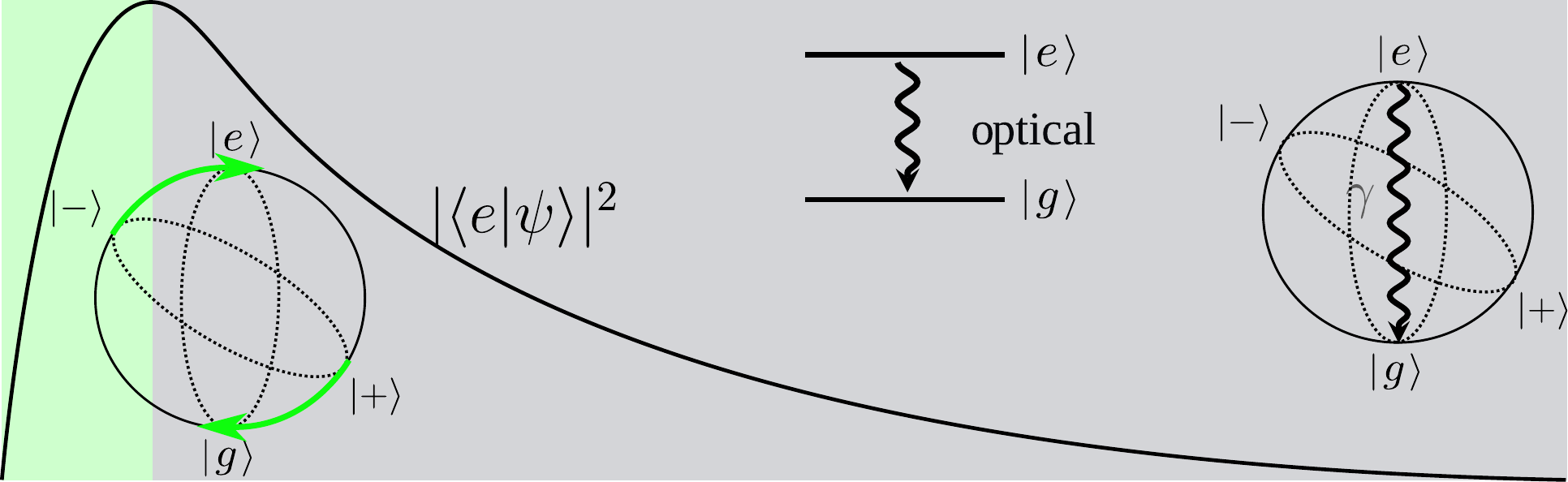}
    \caption{Sketch of the temporal evolution of the population of the population of $|e\rangle$ during the heralding sequence, including Bloch sphere representation.
    }
    \label{heraldingscheme}
\end{figure}

Based on this intuition, we can define two heralding protocols. The first protocol relies solely on the detection of a second optical photon after the activation pulse. While a THz photon can also exit the cavity at times $t>\tau$, the probability of this event is small under the assumption of a strongly dissipative THz cavity, so we rule it out of our calculations for the sake of simplicity. This allows us to write the efficiency of this protocol as

\begin{multline}
    E = P(\text{THz}_{<\tau}|\text{optical}_{>\tau})
    = \frac{P(\text{THz}_{<\tau} \cap \text{optical}_{>\tau})}{P(\text{optical}_{>\tau})} \\
    = \frac{
\int_0^{\tau'} dt' \int_0^{\tau} dt \, \text{Tr}\left[
e^{(\mathcal{L}_0 - \mathcal{K}_\text{opt})(\tau' - t')} \, \mathcal{K}_\text{opt} \,
e^{(\mathcal{L}_0 - \mathcal{K}_\text{opt})t'}
 \mathcal{U} \, e^{(\mathcal{L} - \mathcal{K})(\tau - t)} \, \mathcal{K} \, e^{(\mathcal{L} - \mathcal{K})t} \hat{\rho}_0 
\right]
}{
\int_0^{\tau'} dt' \, \text{Tr}\left[
e^{(\mathcal{L}_0 - \mathcal{K}_\text{opt})(\tau' - t')} \, \mathcal{K}_\text{opt} \,
e^{(\mathcal{L}_0 - \mathcal{K}_\text{opt})t'}
\mathcal{U} \, e^{\mathcal{L} \tau} \hat{\rho}_0 
\right]
},
\end{multline}
where $\text{THz}_{<\tau}$ refers to the emission of a single THz photon during the activation pulse of duration $\tau$, and $\text{optical}_{>\tau}$ to the emission of a single optical photon after the activation and $\hat U$ pulses.

The second protocol, in contrast, heralds the THz photon based on two conditions: the absence of detection of the first optical photon at times $t < \tau$ before the activation pulse is turned off, and the detection of a single optical photon at a later time $t > \tau $.  The associated heralding efficiency (conditional probability) \cite{meyer-scott2017,guilbert2015} of this protocol is given by
\begin{multline}
        \tilde E=P(\text{THz}_{<\tau}| \text{no optical}_{<\tau}\cap \text{optical}_{>\tau})=\frac{ P(\text{THz}_{<\tau} \cap \text{no optical}_{<\tau}\cap \text{optical}_{>\tau})}{ P(\text{no optical}_{<\tau}\cap \text{optical}_{>\tau})}\\
    =\frac{
\int_0^{\tau'} dt' \int_0^{\tau} dt \, \text{Tr}\left[
e^{(\mathcal{L}_0 - \mathcal{K}_\text{opt})(\tau' - t')} \, \mathcal{K}_\text{opt} \,
e^{(\mathcal{L}_0 - \mathcal{K}_\text{opt})t'} \, \mathcal{U} \,
e^{(\mathcal{L} - \mathcal{K} - \mathcal{K}_\text{opt})(\tau - t)} \,
\mathcal{K} \, e^{(\mathcal{L} - \mathcal{K} - \mathcal{K}_\text{opt})t} \hat{\rho}_0
\right]
}{
\int_0^{\tau'} dt' \, \text{Tr}\left[
e^{(\mathcal{L}_0 - \mathcal{K}_\text{opt})(\tau' - t')} \, \mathcal{K}_\text{opt} \,
e^{(\mathcal{L}_0 - \mathcal{K}_\text{opt})t'} \, \mathcal{U} \,
e^{(\mathcal{L} - \mathcal{K}_\text{opt})\tau} \hat{\rho}_0
\right]
}
\end{multline}
 $\tilde E$ corresponds to the ideal heralding protocol, but it is much more challenging to implement, since measuring ``$ \text{no early optical}$" requires extremely high detection efficiencies. In contrast, $E$ is an alternative easier to implement, which, however, does not rule out multiphoton emission of THz photons.

\section{Efficient computation of time integrals of multi-time correlators via Liouvillian spectral decomposition}
In this section, we outline efficient vectorized methods to compute the time integrals of multi-time correlators---quantities that are central for many of the results presented in this paper---in terms of matrix multiplications involving only eigenvectors and eigenvalues of the Liouvillian.  

\subsection{Computing right and left eigenvectors matrices}
We define the $i$-th right and left eigenvectors of the Liouvillian as
\begin{eqnarray}
        \mathcal{L} |r_i\rangle &\equiv& \mathcal{L} \mathbf{r}_i^T = \lambda_i \mathbf{r}_i^T, \label{eq:right-eig} \\
        \langle l_i|\mathcal L & \equiv &\mathbf{l}_i \mathcal{L} = \lambda_i \mathbf{l}_i,\label{eq:left-eig}
\end{eqnarray}
where $\mathbf{r}_i$ and $\mathbf{l}_i$ are row vectors, i.e. $1\times N$ matrices $\mathbf{r_i} = \begin{pmatrix}
        r_{i,1} , r_{i,2} , \ldots , r_{i,N} 
    \end{pmatrix}$, $\mathbf{l_i} = \begin{pmatrix}
        l_{i,1} , l_{i,2} , \ldots , l_{i,N} 
    \end{pmatrix}$
with $N=h^2$, where $h$ is the dimension of the Hilbert space. Each of these objects are to be understood as $h \times h$ matrices arranged in vectorized form. In bra-ket notation, $\ket{r_i}$ is therefore represented as a column vector, $\mathbf{r_i}^T$. The whole set of eigenvectors can be efficiently represented as $N\times N$ eigenvector matrices 
\begin{equation}
    R \equiv \begin{pmatrix}
        \mathbf{r}_1\\ \mathbf{r}_2 \\ \vdots \\ \mathbf{r}_N
    \end{pmatrix}, \quad 
        L \equiv \begin{pmatrix}
        \mathbf{l}_1\\ \mathbf{l}_2 \\ \vdots \\ \mathbf{l}_N
    \end{pmatrix},
\end{equation}
with the corresponding column vector, i.e. $N\times 1$ matrix of eigenvalues
\begin{equation}
        \boldsymbol{\lambda} \equiv \begin{pmatrix}
        \lambda_1\\ \lambda_2 \\ \vdots \\ \lambda_N
    \end{pmatrix}.
\end{equation}
The normalization condition $\mathbf{l}_i \cdot  \mathbf{r}_j = \delta_{ij}$ can then simply be expressed as
\begin{equation}
    L \cdot R^T = \mathbb{I} \label{eq:orthonorm}.
\end{equation}
In order to compute an orthogonalized set, a more robust method than solving the eigenvalue equations \eqref{eq:right-eig} and \eqref{eq:left-eig} independently and then normalizing, is to first solve Eq.~\eqref{eq:right-eig} to obtain $R$. The matrix $L$ of left eigenvectors can then be found by enforcing Eq.~\eqref{eq:orthonorm}, which amounts to computing the inverse of the matrix of right eigenvectors
\begin{eqnarray}
    L = (R^T)^{-1}.
\end{eqnarray}
Working with the eigenvector matrices directly will give us massive computational advantage, since will allow us to use vectorized operations for the calculation of Liouvillian evolution. This efficiency becomes especially important when computing more complex quantities, such as integrals of multi-time correlation operators, as we will demonstrate in the next sections. These derivations will make extensive use of the following spectral decomposition of Liouvillian evolution
\begin{equation}
    e^{\mathcal L t} = \sum_i e^{\lambda_i t}|r_i\rangle \langle l_i|. \label{eq:spectral_decomposition}
\end{equation}

\subsection{Calculation of single-photon probability \( P(\tau) \)}
We are interested in evaluating time-integrated two-time correlation functions of the form
\begin{equation}
    P(\tau) = \int_0^\tau \mathrm{Tr} \left[ e^{\mathcal{L}(\tau - t)} \, \mathcal{K} \, e^{\mathcal{L}t} \rho_0 \right] dt,
    \label{eq:ptau-def}
\end{equation}
where \( \rho_0 \) is the initial density matrix, and \( \mathcal{K} \) is a quantum jump superoperator (e.g., representing a photon detection event). We will now adopt the bra-ket notation, in which a density matrix $\rho$ is represented as a ket $\ket{\rho}$. The representation of this object in algebraic (row vector) form will be denoted as $\boldsymbol{\rho}$. The trace operation can then generally be expressed as $\mathrm{Tr}[\rho] = \braket{\mathbb{I}|\rho} = \mathbf{1}\cdot \boldsymbol{\rho}^T$, where $\ket{ \mathbb{I}}$ is the vectorized form of the identity matrix in bra-ket notation and ${\mathbf{1}}$ is its algebraic (row-vector) representation. Using the spectral decomposition of the Liouvillian, Eq.\eqref{eq:spectral_decomposition}, 
we can expand the integrand in Eq.~\eqref{eq:ptau-def} as
\begin{multline}
    P(\tau) = \int_0^\tau \mathrm{Tr} \left[ \sum_{i,j} e^{\lambda_j (\tau - t)} e^{\lambda_i t} \ket{r_j} \bra{l_j} \, \mathcal{K} \, \ket{r_i} \braket{l_i|\rho_0}\right] dt \\
    = \sum_{i,j}
    \underbrace{\braket{\mathbb{I}|r_j}}_{\equiv T_j}
    \underbrace{\left( \int_0^\tau dt \, e^{\lambda_j (\tau - t)} e^{\lambda_i t} \right)
     \bra{l_j} \mathcal{K} \ket{r_i} }_{\equiv A_{ji}}
      \underbrace{\braket{l_i|\rho_0}}_{\equiv C_i}     = T\cdot A \cdot C^T
\end{multline}
where the terms of the sum have been grouped to represent elements of the $N\times N$ matrix $A$, and the $N$-dimensional vectors $T$ and $C$, which we defined as:
\begin{eqnarray}
    T_j &\equiv& \braket{\mathbb{I}|r_j}, \label{eq:Tj}\\
    A_{ji} &\equiv&\left( \int_0^\tau dt \, e^{\lambda_j (\tau - t)} e^{\lambda_i t} \right)
     \bra{l_j} \mathcal{K} \ket{r_i},\\
     C_i &\equiv& \braket{l_i|\rho_0} \label{eq: Ci}.     
\end{eqnarray}
We can further decompose the matrix $A$ as the Hadamard (element-wise) product of two matrices, $A = I(\tau) \odot B$, where
\begin{align}
    I_{ji}(\tau; \boldsymbol{\lambda}) &\equiv \int_0^\tau dt \, e^{\lambda_j (\tau - t)} e^{\lambda_i t}
    = \begin{cases}
        \tau e^{\lambda_i \tau}, & \text{if } \lambda_i = \lambda_j, \\
        \dfrac{e^{\lambda_i \tau} - e^{\lambda_j \tau}}{\lambda_i - \lambda_j}, & \text{otherwise}
    \end{cases}
    \label{eq:Iji} \\
    B_{ji} &\equiv \bra{l_j} \mathcal{K} \ket{r_i}
    \label{eq:Bij}
\end{align}
If we have previously computed the eigenvector matrices $R$ and $L$, the efficiency of the calculation of $P(\tau)$ can  be greatly enhanced  by pre-computing the vectors $T$, $C$ and the matrix $B$ from equations \eqref{eq:Tj}, \eqref{eq: Ci} and \eqref{eq:Bij}, respectively. Using directly the eigenvector matrices, these objects can be written in a compact and computationally convenient form based on matrix multiplications: 
\begin{eqnarray}
    T &=& \mathbf{1}\cdot R^T, \\
    B &=& L \cdot \mathcal{K } \cdot R^T, \\
    C &=& L \cdot \boldsymbol{\rho}_0^T.
\end{eqnarray}
Having pre-computed these quantitites, evaluating $P(\tau)$ at different values of $\tau$ only requires evaluating the $I(\tau)$ matrix according to Eq.~\eqref{eq:Iji}, and then performing the matrix multiplication:
\begin{equation}
    P(\tau) = T\cdot \left( B\odot
I(\tau) \right) \cdot C.
\end{equation}

\subsection{Calculation of heralding efficiencies}

The method outlined above can be extended to compute more complex quantities, such as the heralding efficiencies \( E(\tau',\tau) \) and \( \tilde E(\tau',\tau) \), introduced in the previous section. Before reminding the explicit expressions of $E$ and $\tilde E$, let us define some relevant effective Liouvillian operators that will be used in their definition, as well as its eigenvectors:
\begin{equation}
\begin{array}{ll@{\quad}l}
S   \equiv \mathcal{L} - \mathcal{K} - \mathcal{K}_{\mathrm{opt}}, & 
\quad \text{with } S \ket{r_i} = \lambda_i \ket{r_i},  \\
S'  \equiv \mathcal{L}_0 - \mathcal{K}_{\mathrm{opt}}, & 
\quad \text{with } S' \ket{r_i'} = \lambda_i' \ket{r_i'},\\
S'' \equiv \mathcal{L} - \mathcal{K}_\text{opt}, & 
\quad \text{with } S'' \ket{r_i''} = \lambda_i'' \ket{r_i''},\\
S''' \equiv \mathcal{L} - \mathcal{K}& 
\quad \text{with } S''' \ket{r_i'''} = \lambda_i''' \ket{r_i'''},\\
S_0 \equiv \mathcal{L}& 
\quad \text{with } S_0 \ket{r_i^0} = \lambda_i^0 \ket{r_i^0}.
\end{array}
\end{equation}

\subsubsection{Heralding efficiency $\tilde E$}
The heralding efficiency $\tilde E$ is then defined as the ratio between two time-integrated two-time correlation expressions, $  \tilde E = \tilde N/\tilde D$, where
\begin{align}
    \tilde N &= \int_0^{\tau'} dt' \int_0^\tau dt \, \mathrm{Tr}\left[
        e^{S' (\tau'-t')} 
        \mathcal K_\text{opt}
        e^{S' t'} 
         \mathcal U
         e^{S(\tau-t)} \, \mathcal{K} \,
        e^{S t} \rho_0
    \right], \label{eq:Nherald} \\
    \tilde D &= \int_0^{\tau'} dt' \, \mathrm{Tr}\left[
        e^{S'(\tau' - t')} \, \mathcal{K}_\text{opt} \,
        e^{S' t'} 
        \mathcal U 
        e^{S'' \tau}
        \rho_0
    \right]. \label{eq:Dherald}
\end{align}

To evaluate these expressions, we follow the same procedure as in the previous section. Using the spectral decomposition of these Liouvillians and the definition of the integrated exponentials of Eq.~\eqref{eq:Iji}, numerator and denominator become:
\begin{eqnarray}
   \tilde N &=& \sum_{m,n,i,j}
    \underbrace{\braket{\mathbb{I} | r'_m}}_{\equiv T_{m}} \,
I_{mn}(\tau', \boldsymbol{\lambda}')
 \,
    \underbrace{
    \braket{l'_m |\mathcal{K}_{\mathrm{opt}} |r'_{n}}
    }_{
    \equiv B'_{mn}} \,
    \underbrace{\braket{l'_n | \mathcal U | r_j}}_{\equiv Q_{nj}} \,
    I_{ji}(\tau,\boldsymbol{\lambda})\,
    \underbrace{\braket{l_j |\mathcal K | r_i}}
    _{B_{ji}}\,
    \underbrace{\braket{l_i|\rho_0}}_{Y_i}\\
 \tilde   D &=& \sum_{m,n,i}
    \underbrace{\braket{\mathbb{I} | r'_m}}_{\equiv T_{m}} \,
I_{mn}(\tau', \boldsymbol{\lambda}')
 \,
    \underbrace{
    \braket{l'_m |\mathcal{K}_{\mathrm{opt}} |r'_{n}}
    }_{
    \equiv B'_{mn}} \,
    \underbrace{\braket{l'_n | \mathcal U | r''_i}}_{\equiv Q'_{ni}} \,
  e^{\lambda''_i \tau}\,
    \underbrace{\braket{l''_i|\rho_0}}_{Y''_i}\\
\end{eqnarray}
where each term is defined as:
\begin{eqnarray}
    T_m &\equiv&  \braket{\mathbb{I} | r'_m}, \\
    B'_{mn} &\equiv&  \braket{l'_m |\mathcal{K}_{\mathrm{opt}} |r'_{n}}, \\
    B_{ji} &\equiv&  \braket{l_j |\mathcal{K} | r_i}, \\
    Q_{nj} &\equiv& \braket{l'_n | \mathcal U | r_j}, \\
    Q'_{ni} &\equiv& \braket{l'_n | \mathcal U | r''_i}, \\
    Y_i &\equiv& \braket{l_i | \rho_0}, \\
    Y''_i &\equiv& \braket{l''_i | \rho_0}.
\end{eqnarray}

Now we can write these matrices using the eigenvector matrices as in the previous section
\begin{align}
    T &= \mathbf{1} \cdot (R')^T, \\
    B' &= L' \cdot \mathcal{K}_{\mathrm{opt}} \cdot (R')^T, \\
    B &= L \cdot \mathcal{K} \cdot R^T, \\
    Q &= L' \cdot \mathcal U \cdot R^T, \\
    Q' &= L' \cdot \mathcal U \cdot (R'')^T, \\
    Y &= L \cdot \boldsymbol{\rho}_0^T, \\
    Y'' &= L'' \cdot \boldsymbol{\rho}_0^T.
\end{align}

Thus, the full expressions for the numerator and denominator take the following expression in terms of matrix products, which benefit greatly from the pre-computation of the matrices defined above:
\begin{eqnarray}
    \tilde N &=& T\cdot \left( I(\tau';\boldsymbol{\lambda'})\odot B'\right) \cdot Q \cdot \left(I(\tau;\boldsymbol{\lambda})\odot B\right) \cdot Y \\
    \tilde D &=&T\cdot \left( I(\tau';\boldsymbol{\lambda'})\odot B'\right)\cdot Q'\cdot (e^{\boldsymbol{\lambda''}\tau}\odot  Y'')
\end{eqnarray}
where we used the notation $\boldsymbol{\lambda}$ to denote the vector of eigenvalues. 

\subsubsection{Heralding efficiency $E$}
Similarly to the previous case, the heralding efficiency $E$ is defined as a ratio $  E = N/D$, where now
\begin{align}
    N &= \int_0^{\tau'} dt' \int_0^\tau dt \, \mathrm{Tr}\left[
        e^{S' (\tau'-t')} 
        \mathcal K_\text{opt}
        e^{S' t'} 
        \mathcal  U
         e^{S'''(\tau-t)} \, \mathcal{K} \,
        e^{S''' t} \rho_0
    \right], \label{eq:Nherald} \\
    D &= \int_0^{\tau'} dt' \, \mathrm{Tr}\left[
        e^{S'(\tau' - t')} \, \mathcal{K}_\text{opt} \,
        e^{S' t'} 
        \mathcal U 
        e^{S_0 \tau}
        \rho_0
    \right]. \label{eq:Dherald}
\end{align}

The calculation is analogous as before. Defining in this case the slightly different matrices
\begin{align}
    T &= \mathbf{1} \cdot (R')^T, \\
    B' &= L' \cdot \mathcal{K}_{\mathrm{opt}} \cdot (R')^T, \\
    B &= L''' \cdot \mathcal{K} \cdot (R''')^T, \\
    Q &= L' \cdot \mathcal  U \cdot (R''')^T, \\
    Q' &= L' \cdot \mathcal  U \cdot (R^0)^T, \\
    Y &= L''' \cdot \boldsymbol{\rho}_0^T, \\
    Y'' &= L^0 \cdot \boldsymbol{\rho}_0^T.
\end{align}
the result is expressed as
\begin{eqnarray}
    N &=& T\cdot \left( I(\tau';\boldsymbol{\lambda'})\odot B'\right) \cdot Q \cdot \left(I(\tau;\boldsymbol{\lambda'''})\odot B\right) \cdot Y\\
   D &=&T\cdot \left( I(\tau';\boldsymbol{\lambda'})\odot B'\right)\cdot Q'\cdot (e^{\boldsymbol{\lambda^0}\tau}\odot  Y'').
\end{eqnarray}

\subsubsection{Analytical expressions for the Heralding efficiency}

\begin{figure}[tbh]
\includegraphics[width=0.7\linewidth]{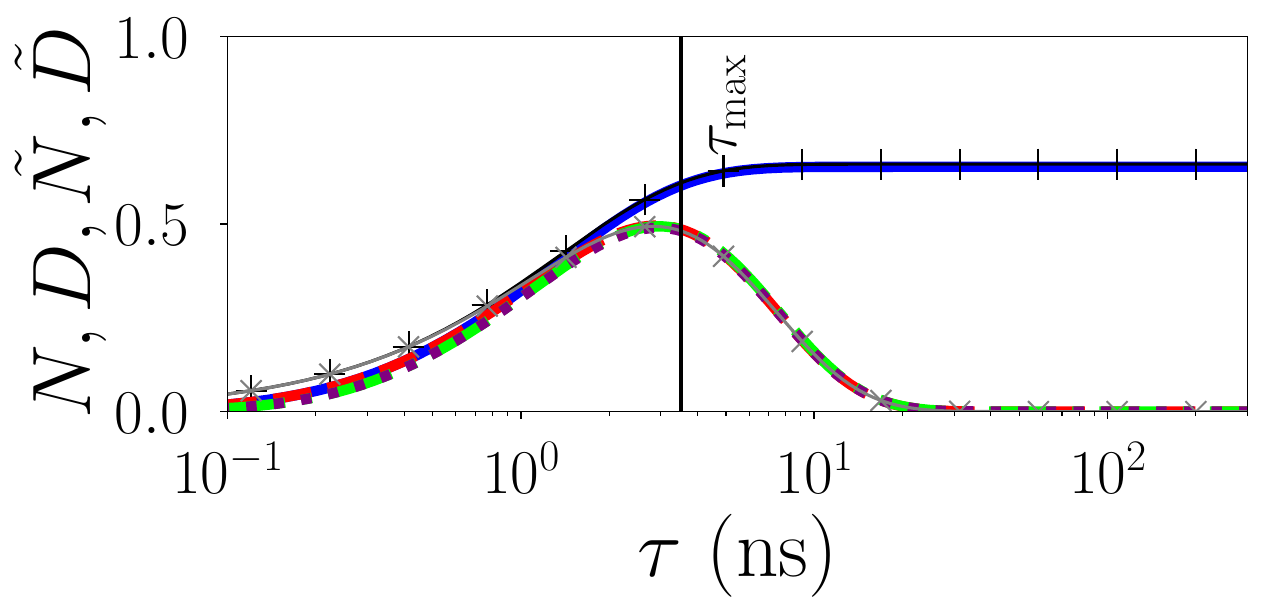}
    \caption{The probabilities $\tilde N$ (dashdotted purple), $N$ (dashed green), $\tilde D$ (dashed red), $D$ (solid blue) and the analytics for $\tilde N, N, \tilde D$ (gray $\times$'s) and $D$ (black $+$'s) as a function of time for the parameter set $\{\gamma,\kappa, \chi,\Omega_R\}/2\pi=\{39.79\times10^{-3},3.5,4,3078\}$ GHz.
    }
    \label{heraldingana}
\end{figure}

Following the procedure in Eq.~(\ref{Panalytics}) we can obtain analytical results for $D$ and $N$. This requires that the model $\mathcal{L}_\text{adb}$ is valid ($\Delta\gg\Omega,\kappa\gg\gamma,g$) and further use that $U\approx1$ and $\mathcal{L}_0=\mathcal{D}(\hat\zeta_+),\mathcal{K}_\text{opt}\hat\rho=\gamma\hat\zeta_+\hat\rho\hat\zeta_-$. We obtain
\begin{eqnarray}
    \tilde N,\tilde D,N&=&-(e^{-\gamma\tau'}-1)\frac{\tilde C}{1-\tilde C}(e^{-\tau\gamma \tilde C}-e^{-\tau\gamma})\\
    D &=&- (e^{-\gamma \tau'}-1)\frac{\tilde C}{1+\tilde C}(e^{-\gamma(1+\tilde C)\tau}+1),
\end{eqnarray}
and
\begin{eqnarray}
    \tilde E&=&1\\
    E&=&\frac{1 + \tilde{C}}{1 - \tilde{C}} \cdot \frac{e^{-\gamma \tau \tilde{C}} - e^{-\gamma \tau}}{1 - e^{-\gamma (1 + \tilde{C}) \tau}}.
\end{eqnarray}
Fig.~\ref{heraldingana} shows the good agreement between analytics and the numerical results for intermediate and large times. Only for small times $<10$ ns a small discrepancy is noticeable. In the short-time regime, all $N$'s and $D$'s converge to the same values, making it impractical to evaluate the efficiencies. However, in the more relevant time frame, where $E$ and $P_1$ reach their maxima, 
the analytical approach remains effective.

\section{Impact of decoherence}
\label{sec:sm-decoherence}
We now consider additional dissipative effects beyond spontaneous emission that may degrade the performance of the source. In particular, we focus here on two mechanisms arising from finite temperature $T$: thermal noise and pure dephasing.
\subsection{Influence of thermal effects}
At finite temperature $T$, the thermal photon occupation number in the cavity will be $\bar n=1/[\exp(\hbar\omega_c\beta)-1]$, where $\beta=1/k_B T$. This arises from the coupling to a thermal bath, described by the following modification of the Lindblad term in the cavity dynamics: 
$\mathcal{D}(\hat X)\rightarrow\bar n \mathcal{D}(\hat X^\dagger)+(\bar n+1)\mathcal{D}(\hat X)$. 

\begin{figure}[t]
    \includegraphics[width=0.7\linewidth]{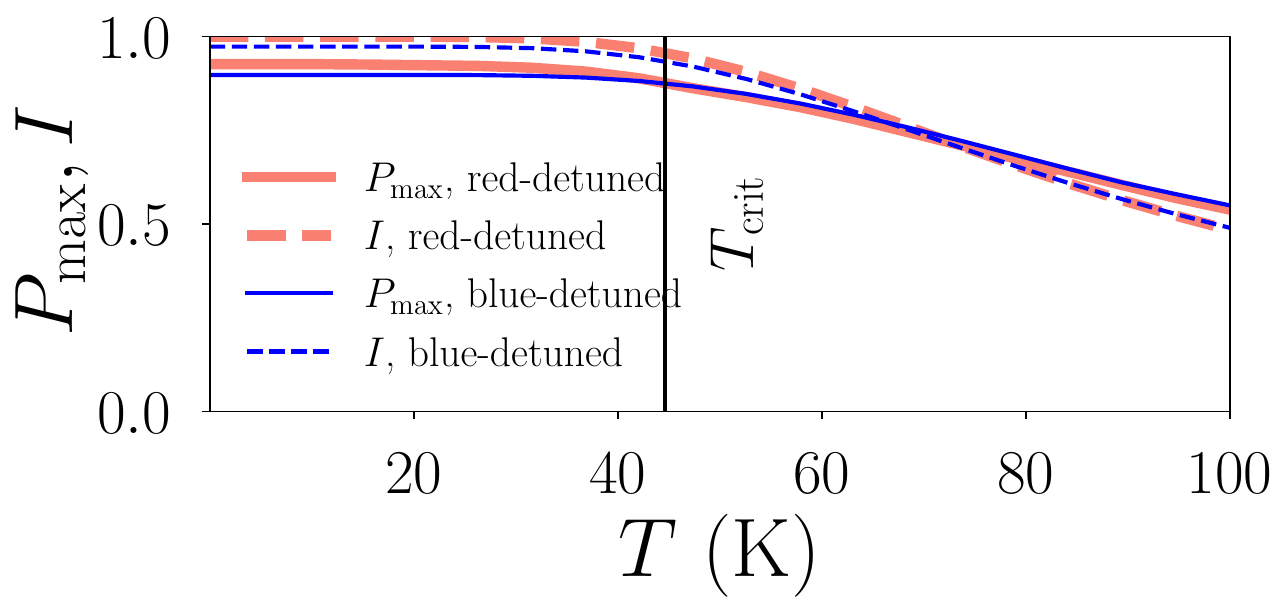}
    \caption{Numerical results for maximum single-photon detection probability $P_\text{max}$ (solid) and indistinguishability $I$ (dashed) as a function the temperature $T$ for the parameter set $\{\chi,\kappa,\gamma,\Omega,\Omega_R\}/2\pi=\{12.5,0.974,39.78\times10^{-3},0.2,5116\}$ GHz for $\Delta>0$ (blue) and $\Delta<0$ (red). 
    The black vertical lines represent the critical value $T_\text{crit}$, where the thermal noise is estimated to begin to dominate.} 
    \label{temperature}
\end{figure}

The most important consequence is that we must now consider that the initial state of the cavity is the thermal state $\hat\rho_T\equiv\sum_{n=0}^\infty \exp(-n\hbar\omega_c\beta)|n\rangle\langle n|/[1-\exp(-\hbar\omega_c\beta)]$. Thermal effects on the population of the emitter and free-space emission can be neglected due to the high visible frequencies yielding negligible thermal occupation numbers. Figure~\ref{temperature} shows the effect of thermal noise on the efficiency, $P_\text{max}$, and the indistinguishability $I$. We can pinpoint a critical temperature below which the dynamics is not affected by thermal occupation and above which there is a sharp decline of both $P_\text{max}$ and $I$ due to the rate associated with the thermal noise $\kappa\bar n$ becoming important, i.e. at least $10$\% of the spontaneous emission rate $\gamma$.
We can estimate the critical temperature $T_\text{crit}=\hbar\omega_c/[k_B\ln(1+10\kappa/{\gamma})]$ by equating these two rates.
For our reference value $\omega_c/2\pi\approx 5.12$ THz the system is largely unaffected up to temperatures of $T=45$ K, enabling optimal operation under standard helium cooling conditions.

\subsection{Influence of pure dephasing}
\begin{figure}
	\includegraphics[width=0.7\linewidth]{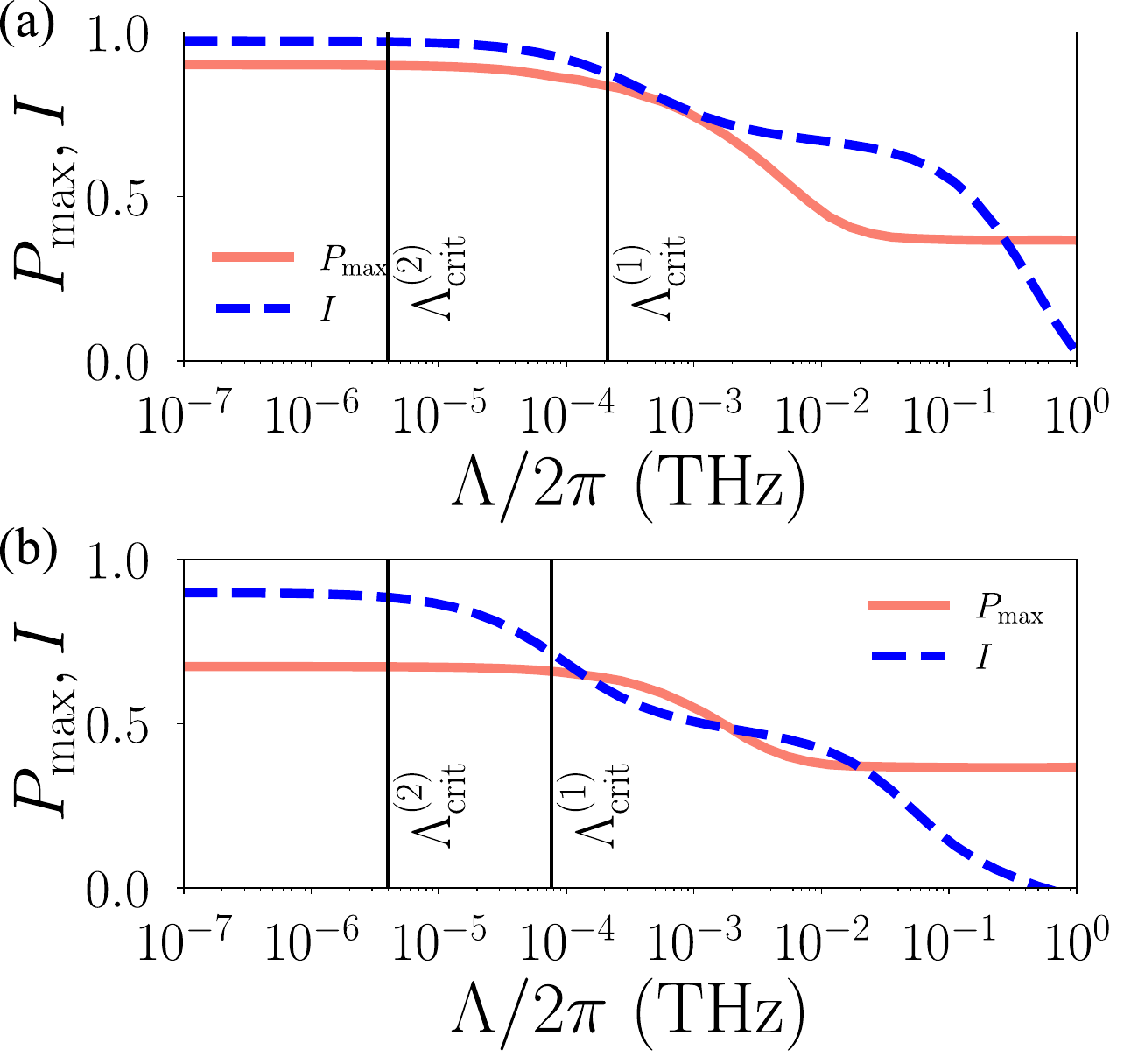}
    \caption{Numerical results for maximum single-photon detection probability $P_\text{max}$ (solid red) and indistinguishability $I$ (dashed blue) as a function of the pure dephasing rate $\Lambda$ for the parameter set (a) $\{\chi,\kappa,\gamma,\Omega,\Omega_R\}/2\pi=\{12.5,0.974,39.78\times10^{-3},0.2,5116\}$ GHz and (b) $\{\chi,\kappa,\gamma,\Omega,\Omega_R\}/2\pi=\{4,3.5,39.78\times10^{-3},0.2,3078\}$ GHz. 
    The black vertical lines represent the critical value $\Lambda_\text{crit}$, where the pure dephasing is estimated to begin to dominate. For the plot with $\omega_c/2\pi=3078$ GHz the indistinguishability is evaluated at $t=0.95$ ns (maximum of $E$ in Fig.~2).} 
    \label{dephasing}
\end{figure}

Another important mechanism that typically limits the performance of single-photon emitters is pure dephasing of the quantum emitter.
In the dressed-state basis, pure dephasing translates into a combination of dephasing, pumping and decay processes, described by the replacement $\frac{\Lambda}{2}\mathcal{D}(\hat\sigma_z)\rightarrow\frac{1}{2}\mathcal{D}(\sqrt{\Lambda_z}\hat\zeta_z-\sqrt{\Lambda_-}\hat\zeta_--\sqrt{\Lambda_+}\hat\zeta_+)$,
where $\Lambda_\pm=4\Lambda c^2s^2$ and $\Lambda_z=\Lambda (s^2-c^2)^2$, respectively. The impact of pure dephasing in the performance of the source is shown in Fig.~\ref{dephasing}, where one sees that efficiency and indistinguishability are impacted at different thresholds (especially if $g\ll\kappa$). In particular, indistinguishability $I$ is more sensitive and begins to degrade at lower values of $\Lambda$ than the efficiency $P_\text{max}$.

The lower impact on efficiency can be explained by decomposing the pure-dephasing Lindblad term into three separate Lindblad channels (under a rotating-wave approximation) and noting the fact that the $\mathcal{D}(\hat\zeta_z)$-term, when added to $\mathcal{L}_\text{adb}$, has no effect on $P_1(\tau)$ as given by Eq.~(1).
Defining a critical dephasing rate for $P_1$ as the point where the remaining pure dephasing terms reach $\Lambda_\pm=0.0081\gamma_+$ yields $\Lambda^{(1)}_\text{crit}=0.0081\gamma c^4/(4c^2s^2)$. Here, the prefactor is motivated by the prior observation that the $\gamma_-$-term started to matter once $h\geq0.3$, corresponding to $\gamma_-/\gamma_+=h^4=0.0081$.

In contrast, the indistinguishability $I$ is directly affected by the $\mathcal{D}(\hat\zeta_z)$-term, which is the dominant one in the off-resonant limit. As a result, the adverse effect on the indistinguishability becomes important at much lower dephasing rates than those affecting the single-photon probability $P_1$. We define a second critical point where pure dephasing rate becomes comparable to the repumping rate $\gamma_+$ and influences the indistinguishability as $\Lambda^{(2)}_\text{crit}=0.1\gamma c^4/(s^2-c^2)^2$. 
For the parameters considered here, this means that dephasing rates should be below $0.2$ GHz to preserve single-photon probability, and below $4$ MHz to maintain high indistinguishability.

\section{Comparison with spontaneous parametric down-conversion sources}
\label{sec:sm-comparison}

The main difference between our proposal and SPDCs lie in their fundamental nature: while here we propose a deterministic source, 
SPDC-based sources are intrinsically probabilistic and generate photon pairs with a distribution over photon-number states. Consequently, detection of an idler photon does not guarantee single-photon emission in the signal mode, and multi-photon contributions are unavoidable~\cite{esmann2024}. 
The probability of generating a single photon per pulse is inherently limited to at most 25\% (two-mode squeezed state in the limit of unity average photon number). That probability can, theoretically, be improved up to \%100 using resource-demanding filtering and multiplexing techniques~\cite{zhang2021,christ2012}. 
Experimentally, reported single-photon probabilities are typically on the order of 4–5\% without multiplexing \cite{tinsley2016,kaneda2019}, and can reach up to 67\% with multiplexing strategies \cite{kaneda2019}. 

In the THz regime, the only relevant demonstration to our knowledge, Ref. \cite{Leontyev2021}, should, in principle, approach the 25\% limit given their reported parametric gain of approximately 0.9. 
In contrast, our scheme yields single-photon probabilities in the range $P_1 = 65$–$92\%$ without multiplexing. In the red-detuned regime, emission is fundamentally deterministic, while in the blue-detuned regime it is effectively deterministic due to the slower repumping dynamics relative to the emission process.

Regarding heralding efficiency, reported values in SPDC systems range from approximately 20\% in unoptimized configurations \cite{tinsley2016} up to 97.7\% with multiplexing \cite{kaneda2019}. Our scheme achieves comparable effective efficiencies in the absence of multiplexing.

Regarding purity, state-of-the-art deterministic single-photon sources in the optical domain report indistinguishability values in the range of approximately 74\%–99\% \cite{esmann2024,kaneda2019}. In our work, the blue-detuned regime yields values of approximately 68\%–88\%, while the red-detuned regime approaches unity due to suppression of repumping-induced mixing. 

\end{document}